\documentclass[journal]{IEEEtran}

\usepackage{cite}      

\usepackage{graphicx}  
\begin{document}
%
\title{Knowledge Network Approach to Noise Reduction}
%
%
\author{Arturo~Berrones
\thanks{
         This work was partially supported by CONACYT 
under project J45702--A, SEP--PROMEP under project 
PROMEP/103.5/05/372
 and UANL--PAICYT
under project CA1275--06.}
\thanks{A. Berrones is with Posgrado en Ingenier\' \i a 
de Sistemas, Facultad de Ingenier\' \i a Mec\'anica y El\'ectrica, 
Universidad Aut\'onoma de Nuevo Le\'on, AP--111, Cd. Universitaria,
San Nicol\'as de los Garza, NL, M\'exico 66450 
(e--mail: arturo@yalma.fime.uanl.mx)}}
\maketitle

\begin{abstract}
Previous preliminary 
results on the application of
knowledge networks to noise reduction in stationary harmonic 
and weakly chaotic signals
are extended to more general cases.
The formalism gives a novel 
algorithm from which
statistical tests for the identification of deterministic 
behavior in noisy stationary time series can be constructed.
\end{abstract}

\begin{keywords}
Noise reduction, knowledge networks, signal processing, time series
analysis.
\end{keywords}

%
\IEEEpeerreviewmaketitle

\section{Introduction}
%
%
%
%
\PARstart{N}{oise} reduction and identification of underlying deterministic
behavior in signals are fundamental questions in fields like communication \cite{wiener, shannon}
and time series analysis \cite{kantz}. A classical model setup relative to the 
measurement of such signals \cite{wiener, shannon}, considers that each
observation in a sequence $y_{1}, y_{2},..., y_i, ..., y_{T}$
can be decomposed as a sum of a deterministic component and a
random perturbation, 

\begin{eqnarray}\label{model}
y_i = y(t_i) = f(t_i)+\varepsilon (t_i) .
\end{eqnarray}

 The random terms $\varepsilon (t_i)$ are statistically independent from measurement
to measurement and independent of $f$.
Consider a clean signal that can be adequately modeled by a 
linear combination of the form

\begin{eqnarray} \label{combination}
f(t_i)=\sum_{l=1}^{L} a_l \varphi (b_l t_i + c_l)
\end{eqnarray}

\noindent where the $\varphi$'s are members of an orthogonal basis
of functions. The meaning of {\it adequately modeled} in the 
present context refers to the consistency of $f$ with Eq. (\ref{model}),
in the following sense:
if $f$ is approximated through the optimization
of some suitable risk or likelihood function in a finite sample, then the residuals
should behave like independent random variables. Additionally, if the resulting form of
$f$ is expected to be used in a fruitful way for prediction purposes, then $f$ should
have the same consistency also outside the original sample, satisfying a suitable
goodness criterion as well.
In general, 
the specific nature of the functional basis for $f$ is hidden. For instance,
the number of components needed to describe the 
signal, $L$, is usually unknown beforehand. Previous to any
attempt of fitting the data to $f$, the model complexity should be defined.
For the setup given by Equations (\ref{model}) and (\ref{combination}) 
$L$ gives a quantity that measures the model complexity. 

The estimation of $L$
is closely related to the separation of the signal from
the noise. In order to see this, consider the case in which $y(t)$ is stationary and
$\left < y \right > = 0$, where the brackets stand for statistical average. The variance of
$y$ is in this case written as

\begin{eqnarray} \label{varia1}
\left < y^{2} \right >= \left <  
\sum_{l_1 =1}^{L} \sum_{l_2 =1}^{L} a_{l_1}a_{l_2}
\varphi_{l_1}(t_i)\varphi_{l_2}(t_i) \right > \\ \nonumber
+ \left < \varepsilon ^ 2 \right > .
\end{eqnarray}

The model complexity $L$ could be estimated 
from the knowledge of the noise amplitude and some
statistical aspects of the
components of the basis. 

The purpose of the present contribution is to
give a novel method for the estimation of the complexity of
signal models, which in turn introduces a new framework to deal with
noise reduction. The main concern regarding the application
of the formalism is on cases in which the noise is {\it strong}, that is,
with a variance comparable with the corresponding variance of
the clean signal.
The proposed approach is valuable
to the characterization of deterministic signals under strong
stochasticity. In many important fields of application, like analysis
of geophysical data, voice recognition, time series
of economic, ecological or clinic origin, etc., the 
identification of deterministic behavior is difficult
due to the presence of strong additive noise or insufficient sample size.
These difficulties are particularily evident for 
the identification and characterization 
of low dimensional chaotic behavior in noisy time series.
The algorithm introduced here tackle these questions for several
important cases. The procedure is linear, yet it is able to perform
signal analysis tasks that are beyond the capabilities of traditional
linear noise reduction techniques.

\subsection{Knowledge Networks}

The proposed method relies on the notion of a 
knowledge network \cite{maslov, berrones}.
Knowledge networks have been originally motivated from the
study of some particular structures that arise in
economy and biology, like interactions between consumers 
and products in a market or protein -- substrate interactions \cite{maslov, maslov2}.  
A knowledge network is defined as a network in which the nodes are
characterized by $L$ internal degrees of freedom, while their edges
carry scalar products of vectors on two nodes they connect \cite{maslov}.
In order to fix ideas, consider the following knowledge network model of
opinion formation \cite{maslov, berrones}: 
suppose that there exists a database of opinions given by agents on a given set of
products. This database can be seen as a sparse matrix, with holes corresponding to
missing opinions (say, agents that have never been exposed to a given product). In
geometrical words, the preferences of an agent are represented as a vector in an hypothetical
taste space, whose dimension and base vectors are generally unknown. A product is represented
by a similar vector of qualities. An agent's opinion on a given product is assumed to be
proportional to the overlap between preferences and qualities, which can be expressed
by the scalar product between corresponding vectors. Therefore,
products act like a basis, and opinions as agent's coordinates on such a basis. 
Consider a population of $M$ agents interacting with $N$ products.
The two sets of vectors lie in a $L$-dimensional space, ${\bf a}_{n}=(a^{1}, a^{2}, ..., a^{L})$ and 
${\bf b}_{m}=(b^{1}, b^{2}, ..., b^{L})$, where $n=1, 2, ..., N$ and $m=1, 2, ..., M$. 
In this way the overlap $y_{m,n}={\bf b}_m \cdot {\bf a}_n$
represents the opinion of agent ${\bf b}_m$ on product ${\bf a}_n$.
Only the overlaps 
$y_{m,n}={\bf b}_m \cdot {\bf a}_n$
can be directly observable. The issue is then to reconstruct the hidden quantities from a known fraction of the scalar products.
For the case in which $L$ is known, Maslov and Zhang \cite{maslov} have shown the existence of thresholds for the fraction $p$ of known overlaps, above which is possible to reconstruct at different extents the missing information. Bagnoli, Berrones and Franci \cite{berrones}, have generalized the study of Maslov and Zhang to the case in which the dimensionality $L$ is unknown. The present work mainly relies on this last approach, so a brief summary of the results of Bagnoli, Berrones and Franci is now presented.

Suposse that the components of $\bf{b}_m$ and $\bf{a}_n$ are random variables distributed according to 

\begin{eqnarray} \label{distro}
P(a_{n}^{l}, b_{m}^{l})=P_{n, l}(a) P_{m, l}(b),
\end{eqnarray}

\noindent
and define $\left < h \right>$ as the average, computed in the thermodynamic limit, over $P(a_{n}^{l}, b_{m}^{l})$ of an arbitrary function $h(a_{n}^{l}, b_{m}^{l})$. For a set of hidden components distributed according to Eq. (\ref{distro}), the $y$'s are uncorrelated in the thermodynamic limit. However, correlations arise because $L$ is finite.

In order to kept the expressions simple, it is assumed that 
$ \left <a_{n}^{l} \right > = \left < b_{n}^{l} \right >=0$. 
Averaging over the distribution (\ref{distro}) the variance of the overlaps is written as

\begin{eqnarray} \label{varia}
\left < y^{2} \right >=L \left < a^{2} \right > \left < b^{2} \right >.
\end{eqnarray}

 For this model setup, Bagnoli, Berrones and Franci \cite{berrones} have shown 
that any overlap can be expressed in terms of a weighted average of other overlaps,

\begin{eqnarray}
y_{m,n} = \frac{L}{M-1} \sum_{i=1}^{M} C_{m,i} y_{i,n} + \epsilon_{L, M, N}, 
\quad i \neq m,
\end{eqnarray}

\noindent where $C_{i, j}$ is the correlation among $y_i$ and $y_j$, 
specifically, the correlation
calculated over
the expressed opinions of agents $i$ and $j$ on different products. This
correlation asymptotically goes to the overlap between the corresponding vectors
of agents tastes. The hidden quantity $L$ can be extracted by fitting the 
proportionality factor $\frac{L}{M-1}$.

The error term $\epsilon$ is at first order given by

\begin{eqnarray} \label{epsilon}
\epsilon \sim \sqrt{\left < a^2 \right >\left < b^2 \right >}L^{3/2}
\frac{\sqrt{M}+\sqrt{N}}{\sqrt{MN}},
\end{eqnarray}

 An aspect of this formalism that is important for applications
is that there is no necessity to have a fully connected opinion matrix. 
The results are extended to sparse datasets simply by the redefinition
of the parameters $M$ and $N$ like functions of the pair $(m,n)$. In this
way $M_n$ represents the available number of opinions over product
$n$ given by any agent and $N_m$ is the number of opinions expressed by
agent $m$ regarding any product \cite{berrones}.

\subsection{Knowledge Networks and Signal Models}

As already pointed out in \cite{encBerrones},
a knowledge network framework
for signals as those described by Eqs. (\ref{model}) and (\ref{combination})
can be built for certain classes of stationary signals.
The essential point is the assumption that a distribution for
the components of the signal model exists, analogous to distribution
(\ref{distro}). If $N$ time ordered
subsamples of size $M$ are extracted from 
the observed sequence $y_{1}, y_{2},..., y_i, ..., y_{T}$, we refer to
$y_{m, n}$ as the measured value at time $m$ in subsample $n$, with $n=1, 2, ..., N$
and $m=1, 2, ..., M$. 
The distribution of the components 
of $y_{m, n}$ is assumed to be

\begin{eqnarray} \label{distro2}
P(a_{n,l}, \varphi_{m,n,l})=P_{n, l}(a) P_{m,n,l}(\varphi).
\end{eqnarray}

In order to see how a distribution $P(a_{n,l}, \varphi_{m,n,l})$ can arise for
the problem in hands, note that from Equations (\ref{model}) 
and (\ref{combination}) follows that

\begin{eqnarray} \label{model2}
y_{m,n}=\sum_{l=1}^{L} a_{n,l} \varphi (m b_{n,l} + c_{n,l})+\varepsilon _{m,n}.
\end{eqnarray}

For fixed $L$, the parameters $a_{n,l}$, $b_{n,l}$ and $c_{n,l}$ are chosen to
be optimal in the given sample with respect to some suitable risk 
or likelihood function \cite{haykin}. Due to the noise and to the finite sample size, the 
chosen parameters fluctuate from sample to sample, giving rise to a distribution
of the form $P(a_{n,l}, \varphi_{m,n,l})$. 

 In the next Section a formalism for noise reduction in signals is built
under the assumption (\ref{distro2}). The close connection between the problem
of noise reduction and estimation of model complexity is shown, leading to a new
technique for model complexity estimation in stationary signals. In Section \ref{ex}
the resulting algorithm is numerically tested on several examples, that are relevant
to important potential applications. Final remarks and a brief discussion of future
work is given in Section \ref{conclusion}. 

\section{Noise Reduction by Knowledge Networks} \label{nr}

Consider the following linear transformation of the components

\begin{eqnarray}\label{change}
a_{n,l} \to a_{n,l} - \left < a_l \right > \\ \nonumber
\varphi_{m,n,l} \to \varphi_{m,n,l} - \left < \varphi_{m,l} \right > ,
\end{eqnarray}

\noindent where 

\begin{eqnarray}\label{means}
\left < a_l \right > = \sum_n a_{n,l}P_{n, l}(a) \\ \nonumber
\left < \varphi_{m,l} \right > = \sum_n \varphi_{m,n,l}P_{m,n,l}(\varphi)
\end{eqnarray}

 Introducing the definitions

\begin{eqnarray}
A= \left (
\begin{array}{c c c}
a_{1,1} & ... & a_{N,1}\\
. &  &  .\\
. &  &  . \\
. &  &  . \\
a_{1,L} & ... & a_{N,L}
\end{array} \right ),
\Phi _{n}= \left (
\begin{array}{c c c }
\varphi _{1,1,n} & ... & \varphi _{M,1,n}\\
. &  &  .\\
. &  &  . \\
. &  &  . \\
\varphi _{1,L,n} & ... & \varphi _{M,L,n}\\
\end{array}
\right )
\end{eqnarray}

\noindent and

\begin{eqnarray}
\Gamma = \left (
\begin{array}{c c c}
\varepsilon_{1,1} & ... & \varepsilon _{1,N}\\
. &  &  .\\
. &  &  . \\
. &  &  . \\
\varepsilon _{M,1} & ... & \varepsilon _{M,N}
\end{array} \right ),
\end{eqnarray}

\noindent the model setup given by Eq. (\ref{model2}) can be written in matricial
form as

\begin{eqnarray}
Y=\Phi_n ^{\tau} A + \Gamma .
\end{eqnarray}

 In the limit $N \to \infty$
the operation $AA^{\tau}$ goes to

\begin{eqnarray}\label{oa}
AA^{\tau} = N \left (
\begin{array}{c c c c }
 \left < a_{1}^{2} \right > & 0 & ... &  0 \\
0 & \left < a_{2}^{2} \right > & ... &  0 \\
 . & . &  &  . \\
 . & . &  &  . \\
 . & . &  &  . \\
 0 & 0 & ... &  \left < a_{L}^{2} \right >
\end{array} \right ) .
\end{eqnarray}

 In the same way, in the limit $M \to \infty$

\begin{eqnarray}\label{ophi}
\Phi \Phi ^{\tau} = M \left (
\begin{array}{c c c c }
\left < \varphi_{1}^{2} \right > & 0 & ... &  0 \\
0 & \left <\varphi_{2}^{2} \right > & ... &  0 \\
 . & . &  &  . \\
 . & . &  &  . \\
 . & . &  &  . \\
 0 & 0 & ... & \left < \varphi_{L}^{2} \right >
\end{array} \right ) .
\end{eqnarray}

 The form of the
diagonal elments in Ec. (\ref{ophi}) follows from an
additional ergodicity assumption: the average
$\left <\varphi_{l}^{2} \right >$
can be equivalently taken
over infinitely many finite samples or
over a single sample of infinite length.
For stationary signals the validity of this assumption
is straightforward. 

 Consider the operation 

\begin{eqnarray} \label{anticipation2}
\hat{Y}=\frac{k}{M} C Y , 
\end{eqnarray}

\noindent where $C$ is the correlation matrix of the $y$'s.
It is now shown that if $\left <a_{l}^{2} \right > = \left <a^{2} \right >$ and 
$\left < \varphi_{l}^{2} \right >=\left < \varphi^{2} \right >$,
that is, if the variabilty due to finite sample size,
discrete sampling and noise affect in the same way all of
the components, then $\hat{Y}=Y - \Gamma$ in the limit $N \to \infty$,
$M \to \infty$, using a suitable value for the factor $k$.
The formula (\ref{anticipation2}) is expanded as

\begin{eqnarray} \label{operation}
\hat{Y} = \frac{k}{M} \frac{YY^{\tau}}{\left < y^{2} \right >} Y
&=& \\ \nonumber
\\ \nonumber
\frac{k}{M} \frac{\left[ \Phi^{\tau}A+\Gamma \right ] 
\left[ A^{\tau}\Phi+\Gamma^{\tau} \right ]}
{N\left[ L\right < a^{2}\left > \right < \varphi ^{2}\left > + 
\right < \varepsilon^{2}\left >\right ]} Y 
&=& \\ \nonumber
\\ \nonumber
\frac{k}{M} \frac{\left [\Phi^{\tau}AA^{\tau}\Phi \Phi^{\tau}A + \Gamma \Gamma^{\tau} \Phi^{\tau}A \right ]}
{N\left [ L\right < a^{2}\left >\right < \varphi^{2}\left >+\right < \varepsilon^{2}\left >\right ]} .
\end{eqnarray}

 Introducing the results (\ref{oa}) and (\ref{ophi}) into Eq. (\ref{operation})

\begin{eqnarray} \label{operation2}
\hat{Y}= \frac{k}{M}  
\frac{M \left < a ^ {2} \right > \left < \varphi ^ {2} \right > +
\left < \varepsilon ^ {2} \right >}{L \left < a ^ {2} \right >
\left < \varphi ^ {2} \right > + \left < \varepsilon ^ {2} \right >}
\Phi ^{\tau}  A .
\end{eqnarray}

 The factor $k$ must therefore be chosen as

\begin{eqnarray} \label{k}
k = \frac{M[L\left < a ^ 2 \right >\left < \varphi ^ 2 \right >+\left < \varepsilon ^ 2 \right>]}
{M\left <a ^ 2 \right >\left <\varphi ^ 2 \right >+\left <\varepsilon ^ 2 \right >}
\end{eqnarray}

 The fluctuations of the observable $y(t)$ can be decomposed as

\begin{eqnarray} \label{varia}
\left < y^{2} \right >=L\left < a^{2}\right >\left < \varphi ^{2}\right > + 
\left < \varepsilon ^ 2 \right > .
\end{eqnarray}

 Introducing Eq. (\ref{varia}) into Eq. (\ref{k}), an expression for $L$ in terms
of measurable quantities is found

\begin{eqnarray} \label{L}
L =  \frac{\alpha M [\left < y^{2} \right >- \left < \varepsilon ^{2} \right >]}
{ \left < y^{2} \right >-\alpha \left < \varepsilon ^{2} \right >} ,
\end{eqnarray}

\noindent where $\alpha = \frac{k}{M}$. In order to see how the terms appearing 
at the right in
Eq. (\ref{L}) are measured, consider
the following algorithm for
noise reduction and estimation of the optimum complexity in models for stationary signals.
The anticipation formula in this case reads

\begin{eqnarray} \label{anticipation}
\hat{y} (t_i) = \frac{k}{M} \sum_{h=1, h \neq i}^{M} C(t_h) y(t_i - t_h) .
\end{eqnarray}

 The signal is processed performing the following steps:

i) Calculate the autocorrelation function $C(t)$.

ii) Perform mean squares over a sample of $M$ consecutive points to estimate
the factor $\alpha = \frac{k}{M}$ in Eq. (\ref{anticipation}).

 The mean squares problem can be solved exactly, giving 

\begin{eqnarray}
\alpha = \frac{\sum_{i=1}^{M} y(t_i)
\sum_{\tau=1}^{M} C(t_{\tau}) y(t_i - t_{\tau})}
{\sum_{j = 1}^{M} \sum_{\tau _1 = 1}^{M} C(t _{\tau _1}) y(t_j - t _{\tau _1}) 
\sum_{\tau _2 = 1}^{M} C(t _{\tau _2}) y(t_j - t _{\tau _2})
} \\ \nonumber
i \neq \tau, j \neq \tau _1, j \neq \tau _2,
\end{eqnarray}

\noindent with $M$ less than or equal to one half of the total lenght of the signal. 
The term $\left < \varepsilon ^{2} \right >$ is estimated after the filtering, using the filtered 
data as an approximation of the underlying deterministic signal and performing the
substraction $\left < \varepsilon ^{2} \right >=\left < y ^{2} \right >
-\left < f ^{2} \right >$

iii) By the use of Eq. (\ref{L}), calculate $L$ in terms of observable quantities.

 The steps i) -- iii) define what hereafter is called the Knowledge 
Network Noise Reduction (KNNR) algorithm.

\section{Examples}\label{ex}

The KNNR algorithm is tested on data generated numerically, 
adding at each time step a Gaussian white noise term
$\varepsilon (t)$ 
to a deterministic function $f(t)$.
The simulation of the noise is based on the L'Ecuyer algorithm, which is
known to accomplish adecuate performance with respect to the
main statistical tests, and to produce sequences of random
numbers with lenght $\sim 10^{18}$ \cite{nr}. The noisy data
$y(t)=f(t)+\varepsilon(t)$ enters as input for the KNNR algorithm.
By the use of the Fast Fourier Transform of the
input \cite{nr}, the autocorrelation function is calculated for a maximum
lag equal to one half of the total length of the signal. The steps 
ii) and iii) of the KNNR algorithm are then performed over the second half of
the input.

 The capabilities for noise reduction 
in harmonic and weakly chaotic time series
of the proposed method have
already been discussed in \cite{encBerrones}. 

 The KNNR framework provides a 
characterization
of the signal model complexity in terms of $L$, 
the number of member functions
of a certain orthogonal basis needed to describe the signal, if it
is indeed separable into a deterministic component and a white noise term.
If the necessary assumptions are met, $L$ should converge to a finite value
as the sample size grows.
This fact can be used to identify underliyng deterministic 
behavior. 

 In the next examples the KNNR approach is tested on several chaotic
systems, with and without additive noise, 
and for camparision purposes,
on purely stochastic systems as well. 
The mean value of the signals is substracted before 
they enter as input in the KNNR algorithm. The examples 
with a deterministic part are therefore constructed by

\begin{eqnarray}
y_i=s_i + \varepsilon _i - \left < y \right > ,
\end{eqnarray}

\noindent where $y_i$ is the input and $s_i$ is given by the iteration of
a nonlinear discrete map. Each of the noise terms $\varepsilon _i$, is 
independently drawn from a Gaussian distribution.

 The KNNR algorithm is capable to perform tasks that are beyond
the scope of traditional linear signal processing techniques.
For instance, with large enough sample size, the KNNR algorithm is able
to identify nonlinear behavior in signals whose power spectrum 
is consistent with a correlated stochastic process. This identification
is not possible by classical approaches like the Wiener filter \cite{wiener},
which relies in a clear separation between oscillatory and noise components
in the spectrum. More recent methods, like surrogate data \cite{surrogate}
or nonlinear techiques \cite{barahona}, on the other
hand, do not give a comprehensive framework to deal with noise reduction and 
identification of determinism in a common ground.

\subsection{The Logistic Map}

An archetypal example of a simple
nonlinear system capable of chaotic behavior is given by
the logistic map \cite{ott}

\begin{eqnarray}\label{logmap}
s_i=rs_{i-1}(1-s_{i-1}) .
\end{eqnarray}

 With
a parameter value of $r=3.6$ 
and initial conditions in the interval $(0,1)$,
the map (\ref{logmap}) displays a weakly chaotic behavior, 
close to quasi -- periodic 
motion. As already discussed in \cite{encBerrones}, in this case
$L \sim 2$, indicating that with this low model complexity
is possible to accomplish the separation dictated by Eqs. (\ref{model}) 
and (\ref{combination}). 

 A case with $r=3.7$ 
and the initial condition in the interval $(0,1)$
is analyzed with the
KNNR algorithm. The map is
perturbed by
a Gaussian white noise with a variance of $0.2$, 
essentially the same variance of
the clean signal. 

\begin{figure}[h] 
\vskip0.5cm
\centering{\resizebox{6cm}{!}{\includegraphics{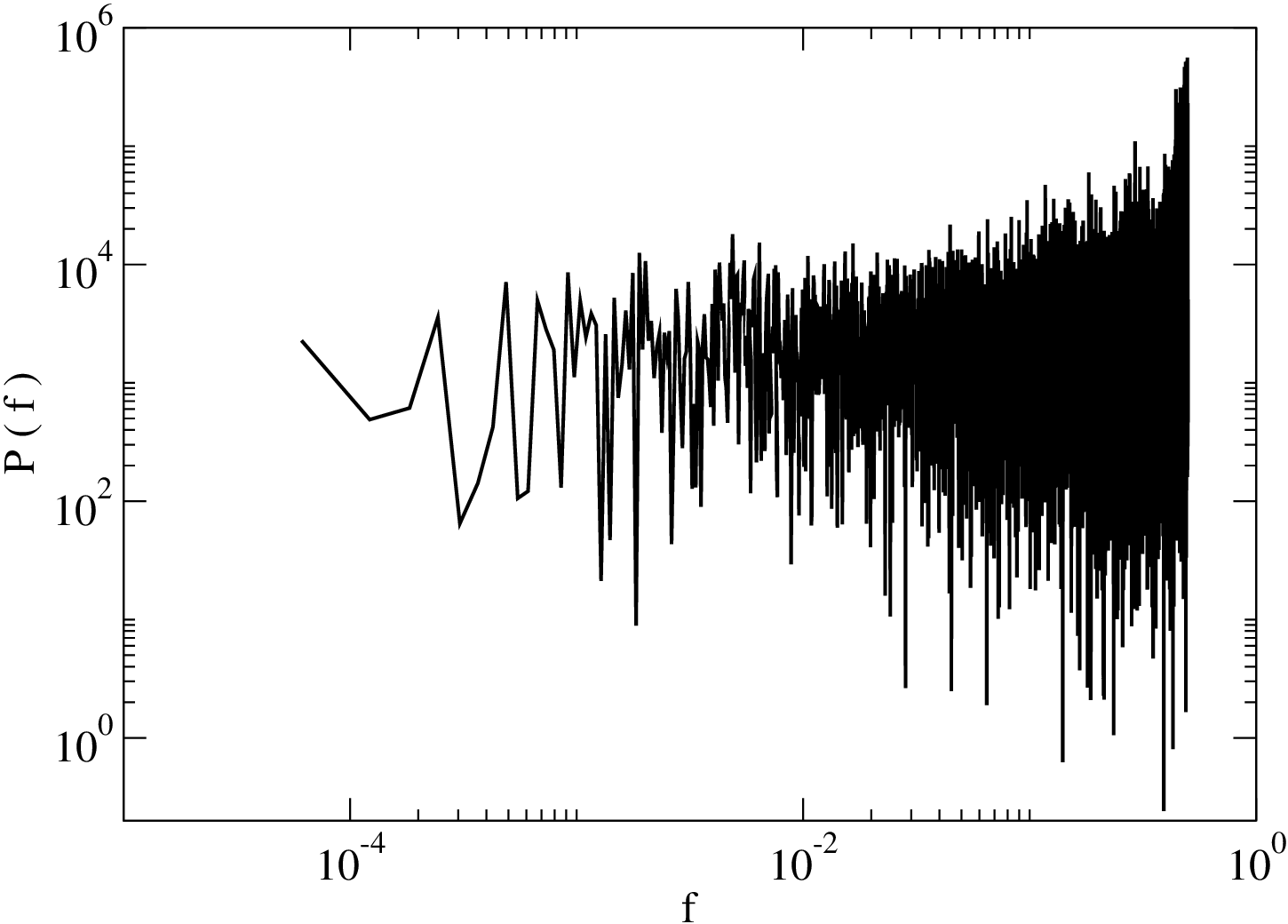}}}
   \caption{\label{LogisticPS} Log--log plot of the power spectrum of the
perturbed logistic map.}
\end{figure}

 The power spectrum taken from a sample of $16384$ points
of the input signal is presented in Fig. \ref{LogisticPS}.
Besides the presence of some relevant peaks at high frequencies,
the spectrum is basically a white noise. 

\begin{figure}[h] 
\vskip0.5cm
\centering{\resizebox{6cm}{!}{\includegraphics{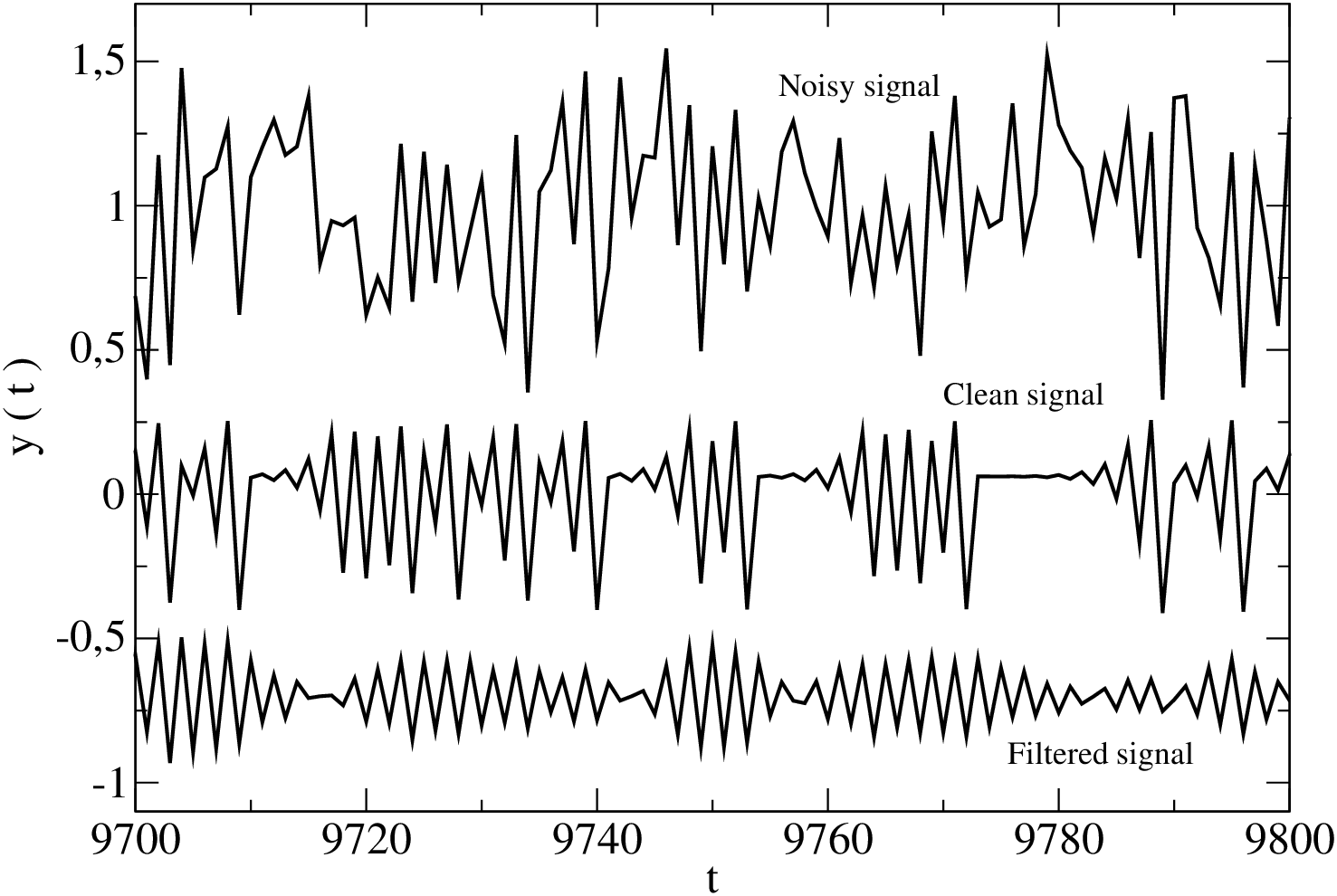}}}
   \caption{\label{logistic} Noise reduction by the KNNR algorithm
for a strongly perturbed logistic map.}
\end{figure}

 Segments of the noisy, clean and filtered time series
are shown in Fig. \ref{logistic}. 
In order to present all the data in the same graph, 
suitable constants have been added to the mean values of the signals.

\begin{figure}[h] 
\vskip0.5cm
\centering{\resizebox{6cm}{!}{\includegraphics{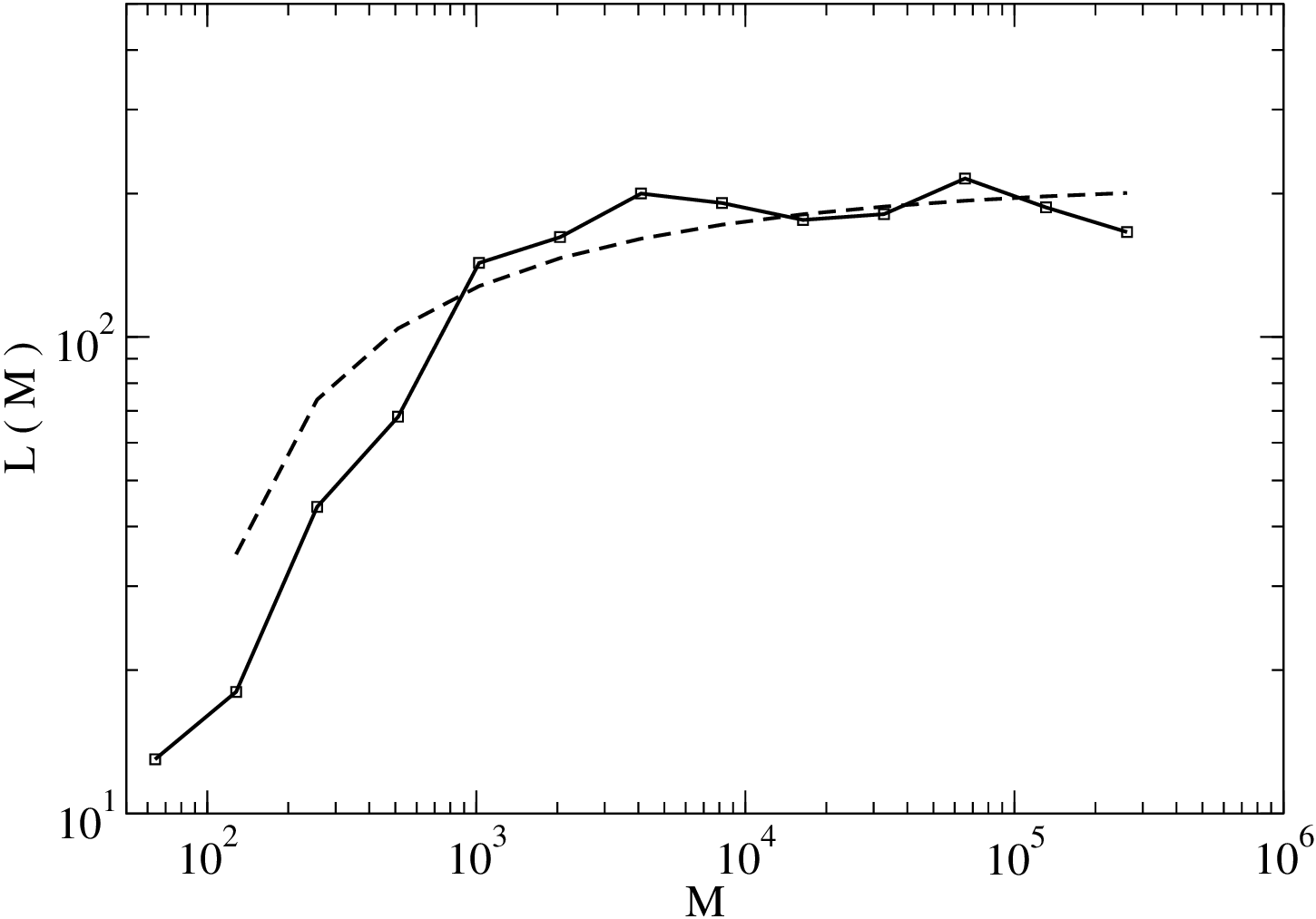}}}
   \caption{\label{convergenceLogistic} Convergence of $L$ for the perturbed
logistic map.}
\end{figure}

In Fig. \ref{convergenceLogistic} the values of $L$ for increasing sample size are plotted.
A mean squares fit of the resulting data is
performed with respect to the formula

\begin{eqnarray} \label{convergence}
L_M = L - a M^{-\frac{1}{2}}, \quad a > 0 .
\end{eqnarray}

 The type of convergence given in Eq. (\ref{convergence}) is suggested by
the first order error term Eq. (\ref{epsilon}), in the anticipation formula of the
original Bagnoli, Berrones and Franci setup.
This behavior of errors is obtained for the case in which the basis components are
independent random variables. The fundamental point in the derivation of 
Eq. (\ref{epsilon}) is that the fluctuations of these components sum in 
accordance to the Central Limit Theorem \cite{berrones}. The numerical results suggest
that for strongly chaotic systems this condition holds. In this example
the number of hidden components converge to a value of order $L \sim 10^2$.

 The convergence of $L$ constitute a basis
for a novel technique of identification of chaos 
and other types of deterministic behavior in time series.
In real world problems, the availability of 
arbitrarily large samples is a rare luxury. The convergence of
$L$ can be however assesed indirectly, through the parameter $\alpha$ that appears
in formula (\ref{L}).
As the sample size grows, the variance terms of Eq. (\ref{L}) tend to be
constant.
In order to have a finite value for $L$, $\alpha$ must be decreasing with $M$. 
Of course, assimptotically $\alpha \propto M^{-1}$. 
The particular way in which this assimptotic behavior is attained is
unknown.
By a smootness assumption
a decreasing behavior of $\alpha$ can be however expected for a 
range of sample sizes.
Note that this claim is consistent with the curve shown in Fig. \ref{convergenceLogistic}.
On the other hand, according to the evidence presented in Subsection \ref{stochastic},
$L$ diverges assimptotically in a linear way with sample size for
linear stochastic processes with finite correlation lengths.
On these grounds, the proposed test for determinism is a standard $F$--test applied to
$log [\alpha (M)]$. 
Consider the model $log (\alpha) = - \beta log (M) + c$, where $\beta$ is a positive
number and $c$ is a real. These parameters are given by 
fitting the linear model to data.
The null hypothesis is that $\beta = 0$.
Numerical results indicate that the proposed test gives a reliable identification 
with input signals of moderate length. 
In this and all of the following examples the $F$-test is performed 
over a set of values of the parameter $\alpha$ calculated through the KNNR algorithm
for noisy signals with sample sizes of $64$, $128$, $256$, $512$, $1024$, $2048$, $4096$, $8192$ 
and $16384$. 

\begin{figure}[h] 
\vskip0.5cm
\centering{\resizebox{6cm}{!}{\includegraphics{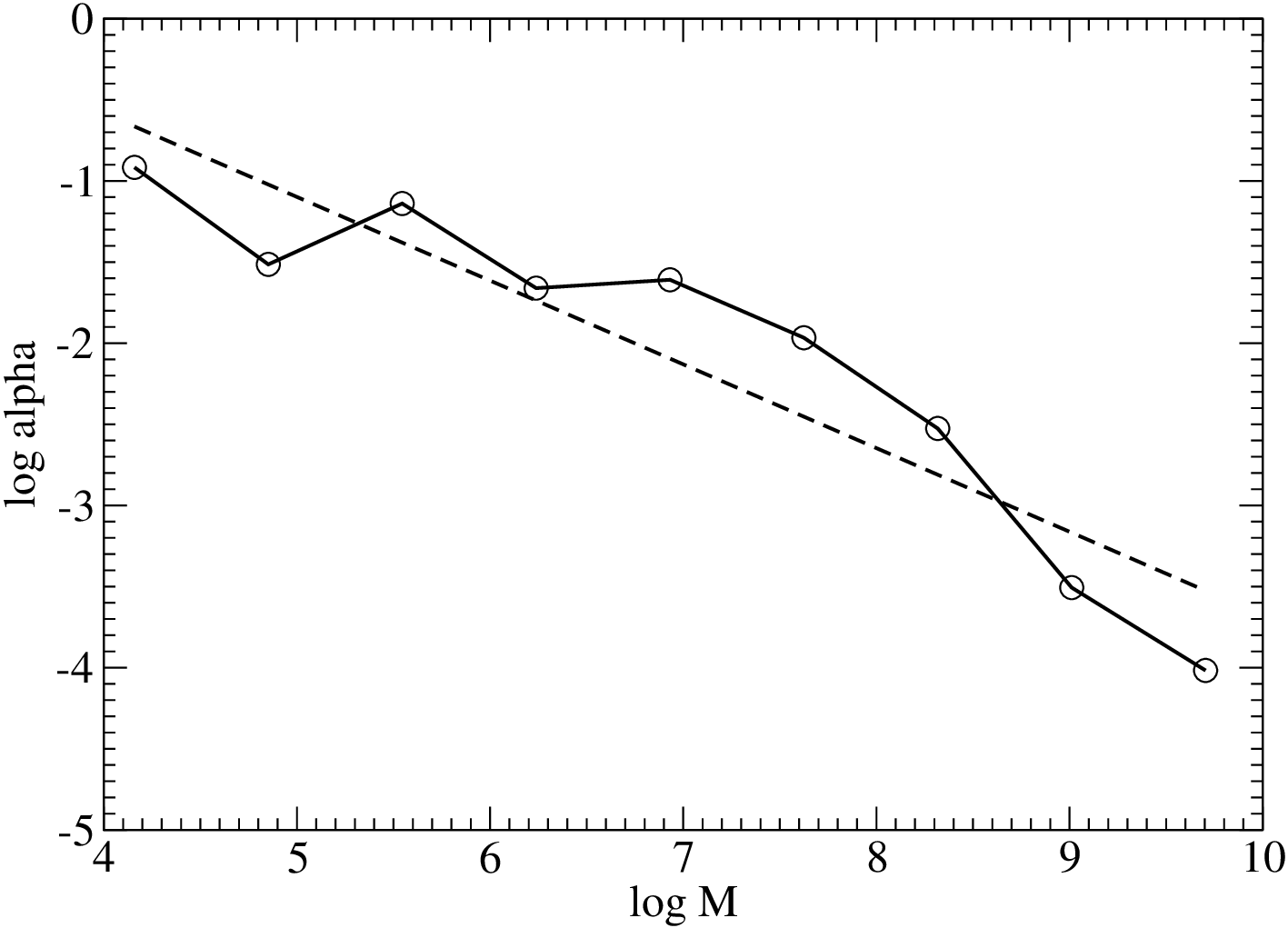}}}
   \caption{\label{alphaLogistic} Behavior of $\alpha$ with increasing $M$
for the noisy logistic map.}
\end{figure}

In Fig. \ref{alphaLogistic} is presented $log (\alpha)$ vs $log (M)$, where $log$
stands for the natural logarithm. 
It turns out 
that $F = 42 >> F_{0.05}(1,7) = 5.59$, so the null hypothesis is clearly rejected at 
a $95 \% $ confidence level.

\subsection{The H\'enon Map}

A famous two dimensional extension of the Logistic Map is the system introduced by 
H\'enon \cite{henon},

\begin{eqnarray} \label{henon}
s_i=a-s_{i-1}^{2}+bx_{i-1}, \\ \nonumber
s_i=x_{i-1}.
\end{eqnarray}

\begin{figure}[h] 
\vskip0.5cm
\centering{\resizebox{6cm}{!}{\includegraphics{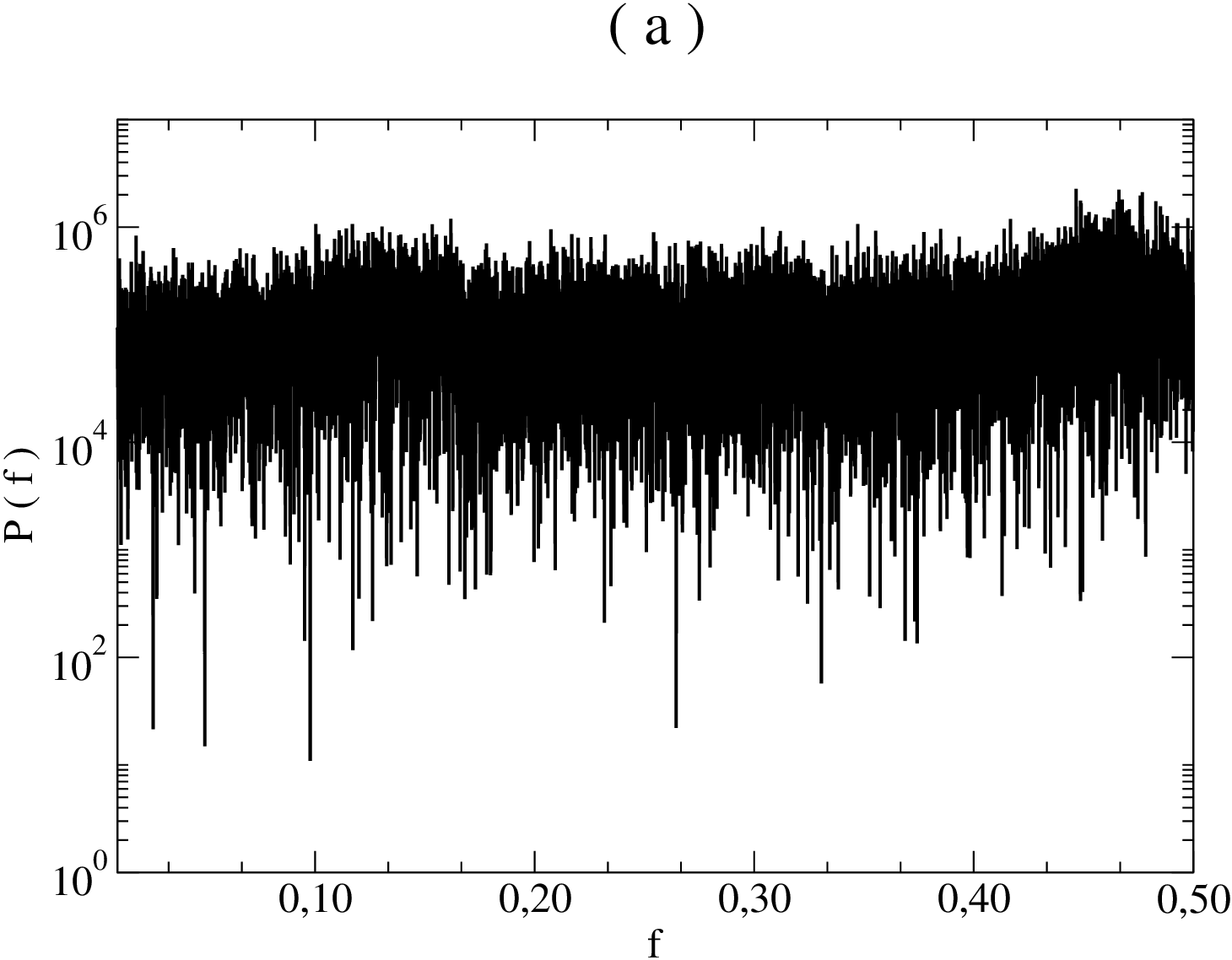}}}
\vskip1.0cm
\centering{\resizebox{6cm}{!}{\includegraphics{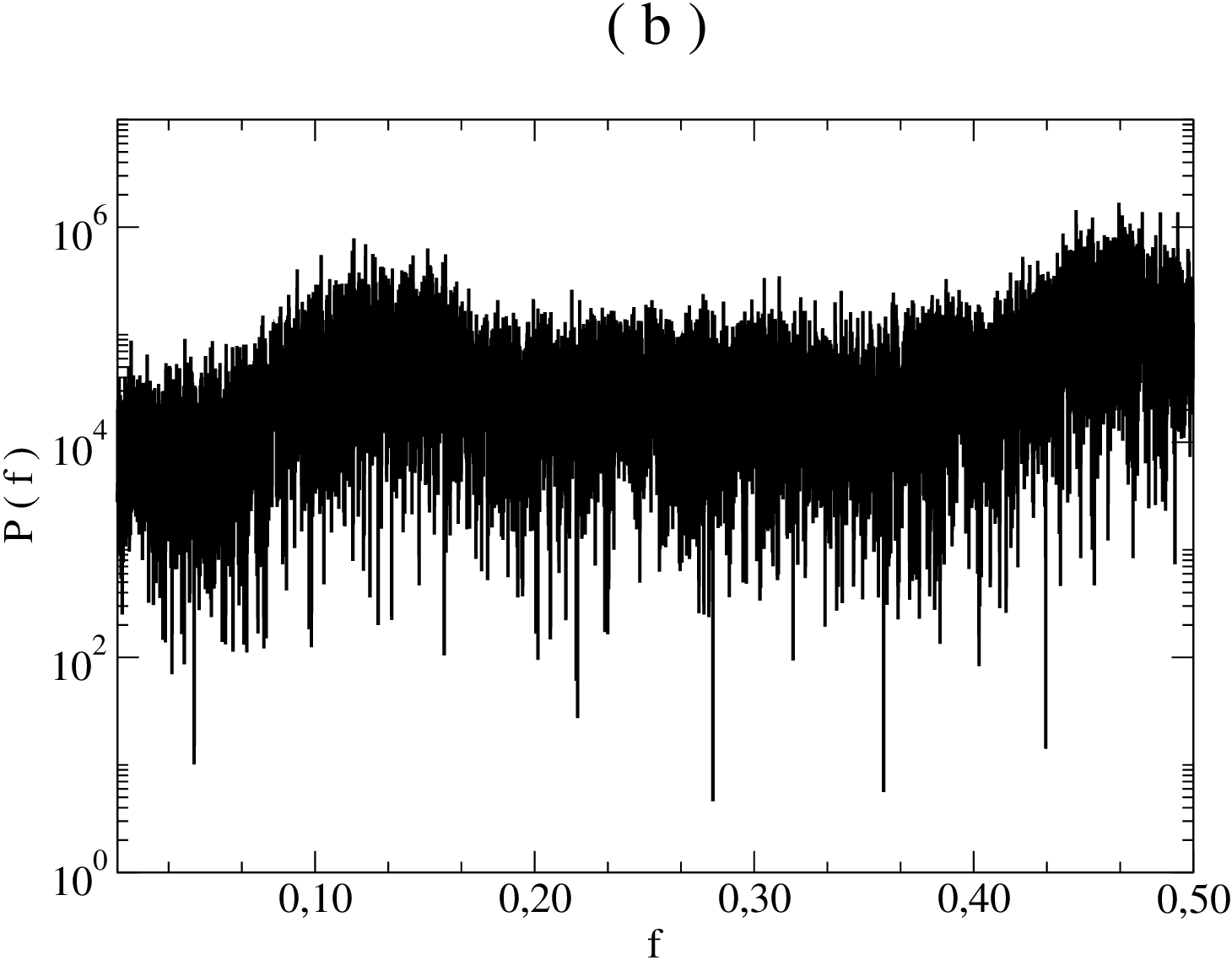}}}
\vskip1.0cm
\centering{\resizebox{6cm}{!}{\includegraphics{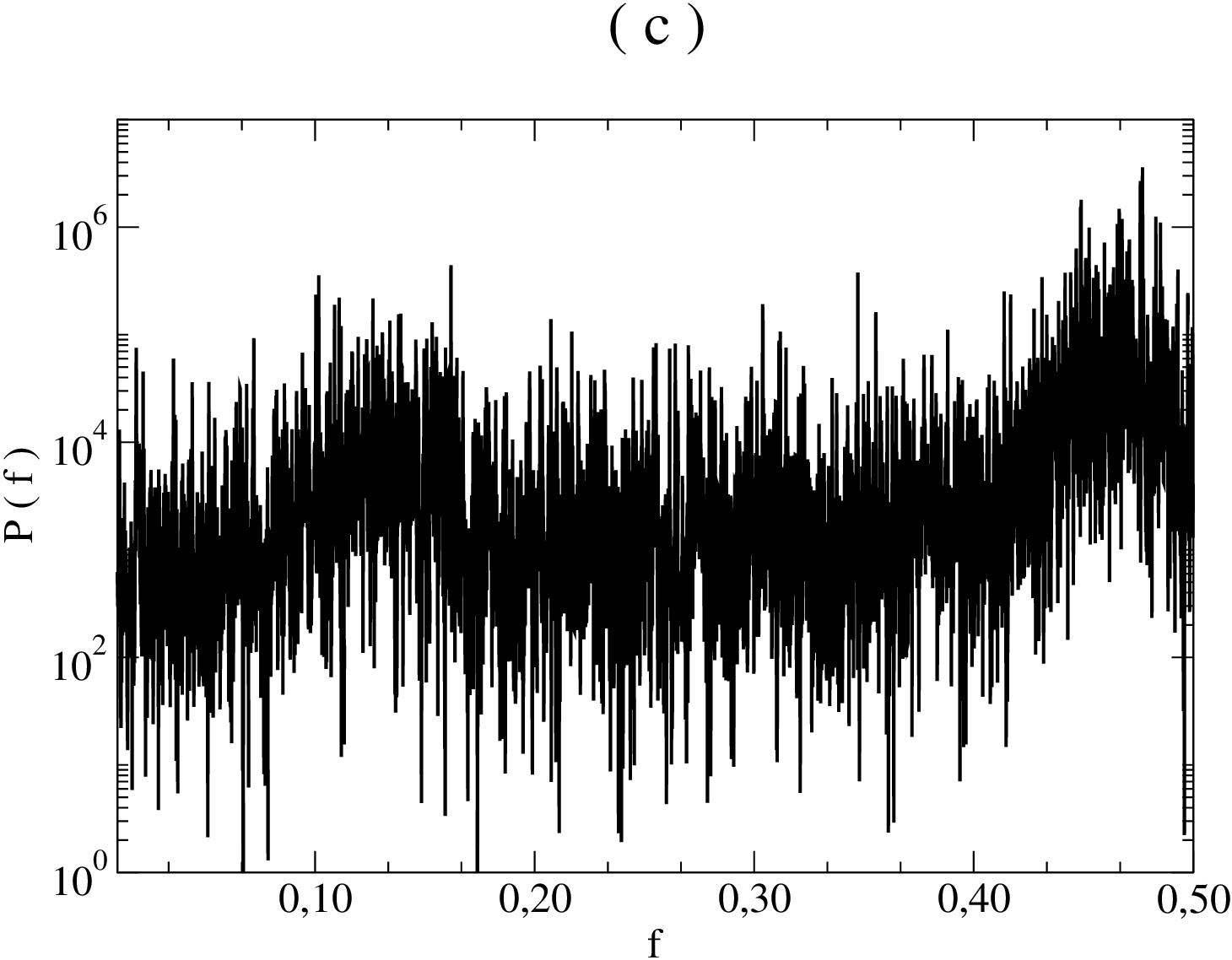}}}
   \caption{\label{figHenon} Semilog plots of the power spectra of a signal
generated by the H\'enon map: (a) noisy case, (b) clean signal, 
(c) filtered signal. }
\end{figure}

The canonical values $a=1.4$, $b=0.3$ are taken. 
The iteration of the map (\ref{henon}) 
is done starting from the initial conditions $s_0 = 0.5$, $x_0 = 0.5$.
The KNNR algorithm is applied to a case in presence of
a noise with variance $1.2$ (the 
variance of the clean signal is $1$).
The knowledge network algorithm performs a satisfactory noise reduction
of the input signal. Fig. \ref{figHenon} shows the power spectrum of the clean, noisy and
filtered signals in semilog scale. 
The input has a length of $16384$ points.
The filtered signal captures the overall shape
of the clean spectrum.

For the noisy H\'enon system the $F$--test again indicates convergence in $L$
at a $95 \% $ confidence level.
It is found that $F = 7.2 > F_{0.05}(1,7) = 5.59$.

\subsection{The Intermittency Map}

In this example the deterministic part of the input is generated
by the iteration of the intermittency map,

\begin{eqnarray}\label{intermittency}
s_i = \beta + s_{i-1} + cs_{i-1}^{m}, \quad 0 < s_{i-1} \leq d \\ \nonumber
= \frac{s_{i-1} - d}{1-d}, \quad d < s_{i-1} < 1  \\ \nonumber
c = \frac{1- \beta - d}{d^{m}} .
\end{eqnarray}

 The map (\ref{intermittency}) is related to several models that arise
in the study of the phenomenon of intermittency found in turbulent fluids \cite{schuster}.
Recently, the map (\ref{intermittency}) has been proposed as a model for the long term
dynamics of packet traffic in telecommunication networks \cite{herramilli}.

 Depending on the parameters, the system (\ref{intermittency}) can display spectral
properties that range from white noise to $1/f$ noise. 

 The values for the parameteres $m$ and $d$ considered here are
$m=2$, $d=0.7$. 
The initial condition is taken as $0.01$.
Two different cases are studied: 

 i) $\beta = 0.05$. 

\begin{figure}[h] 
\vskip0.5cm
\centering{\resizebox{6cm}{!}{\includegraphics{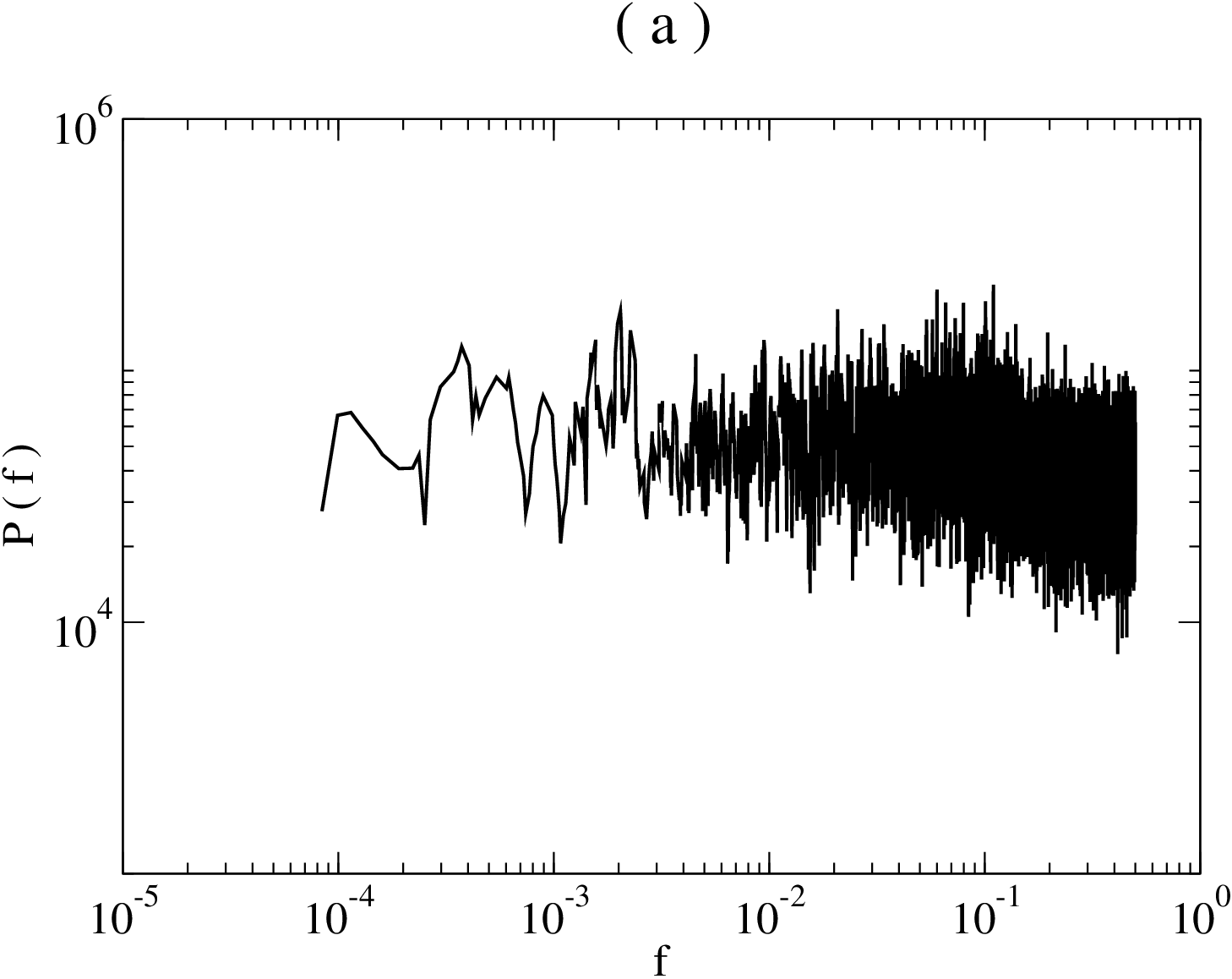}}}
\vskip1.0cm
\centering{\resizebox{6cm}{!}{\includegraphics{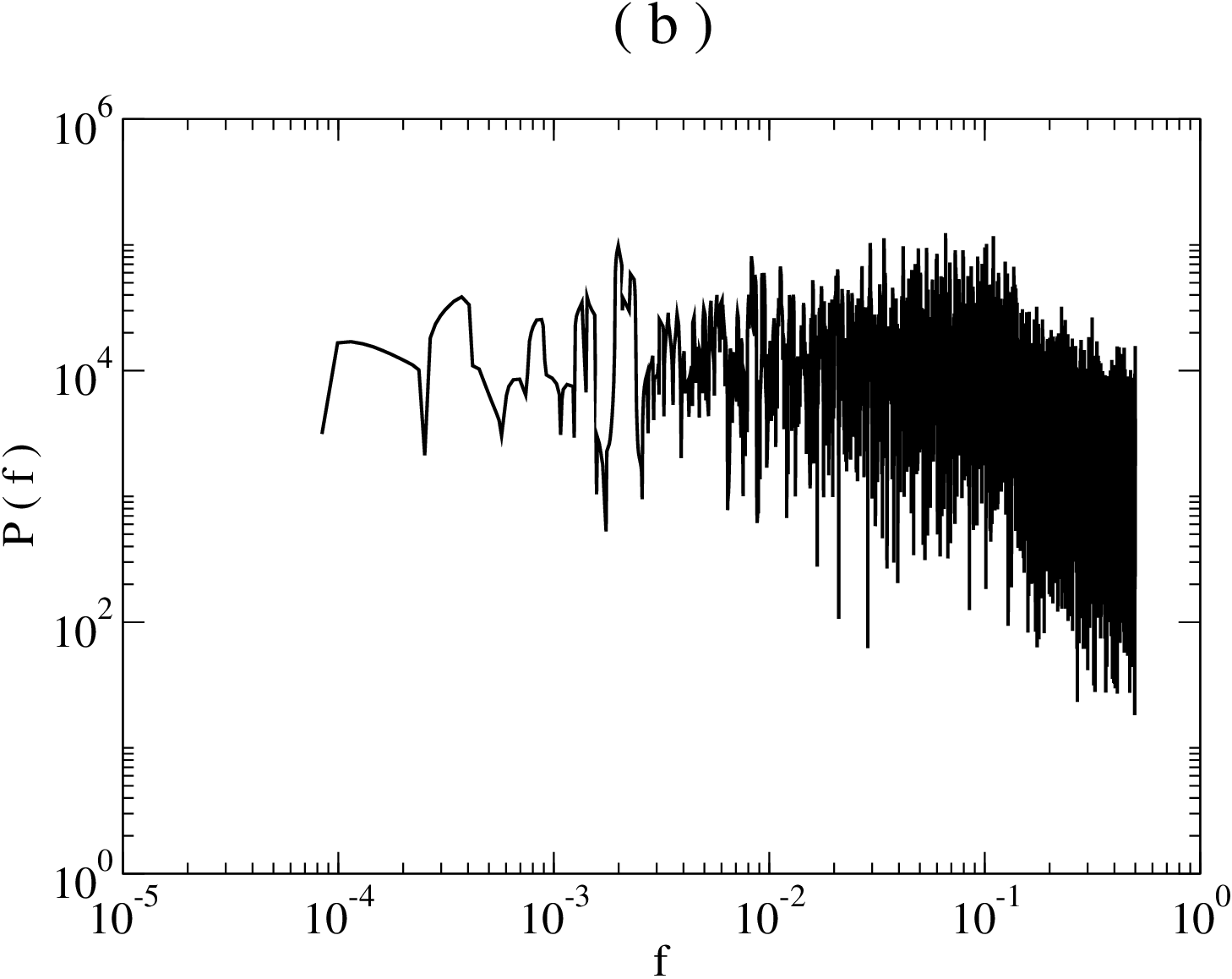}}}
\vskip1.0cm
\centering{\resizebox{6cm}{!}{\includegraphics{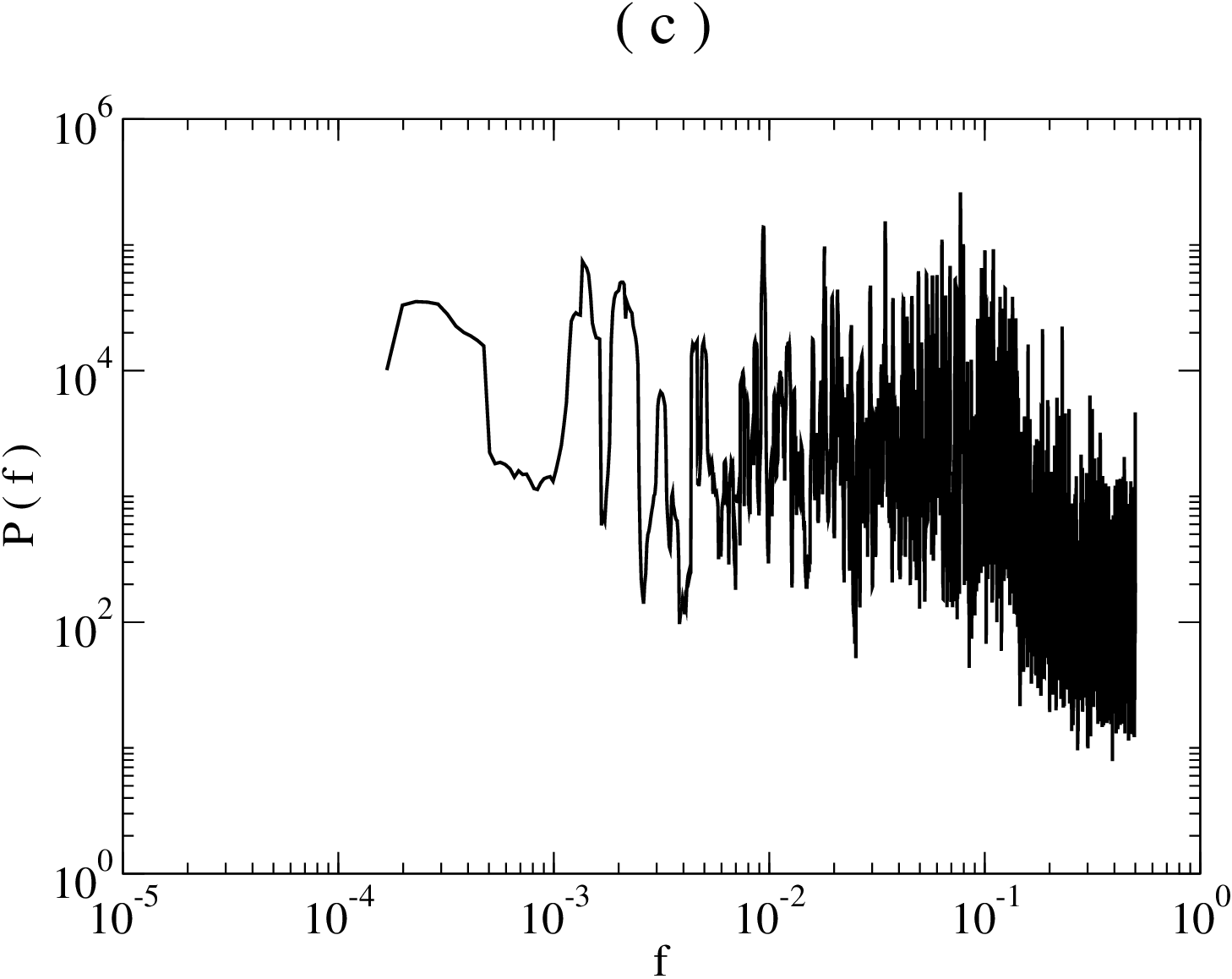}}}
   \caption{\label{intermittency2} Log--log plots of the power spectra of a signal
generated by the intermittency map ($\beta = 0.05$): (a) noisy case, (b) clean signal, 
(c) filtered signal. }
\end{figure}

\begin{figure}[h] 
\vskip0.5cm
\centering{\resizebox{6cm}{!}{\includegraphics{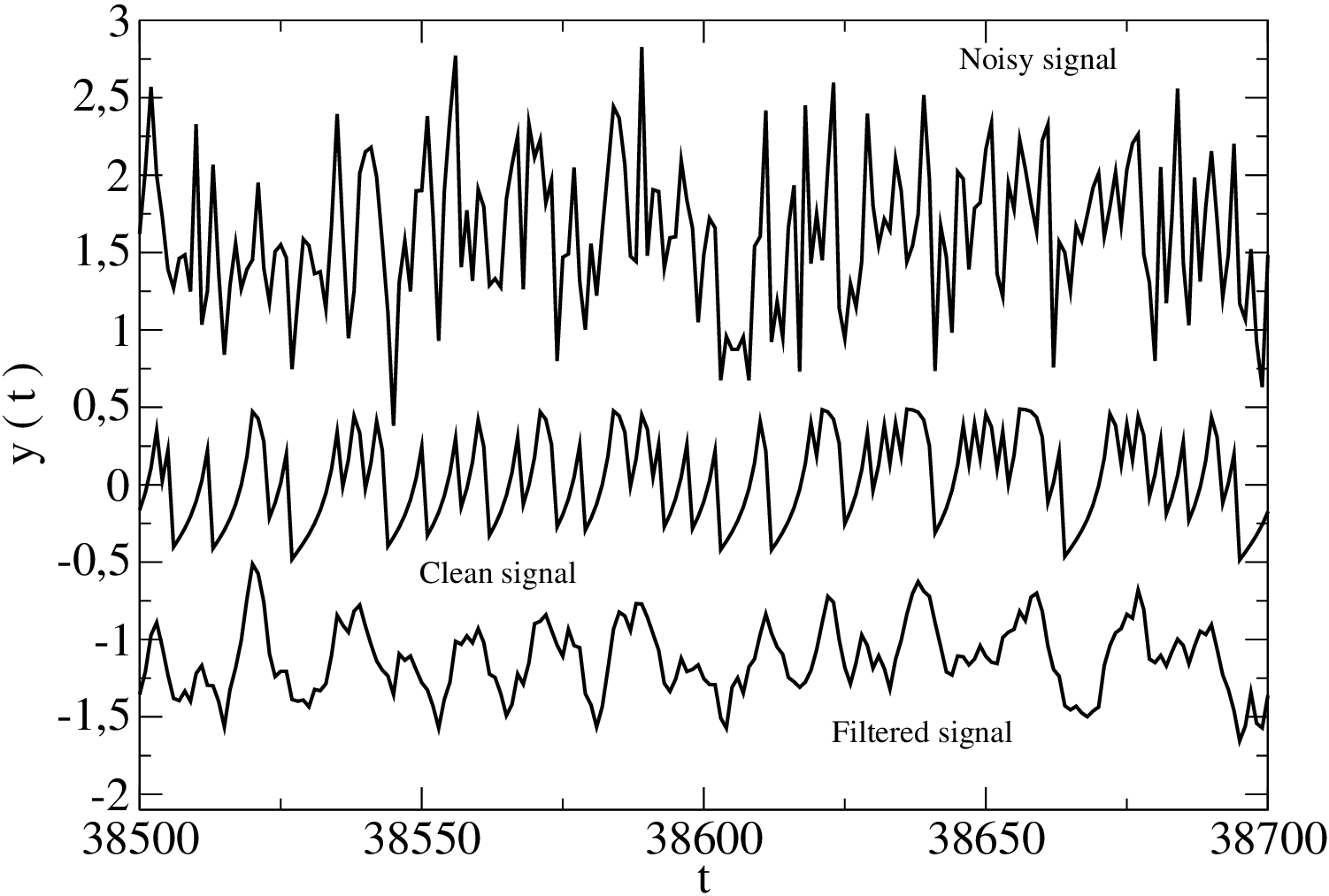}}}
  \caption{\label{intermittency1} Noise reduction for a strongly
perturbed intermittency map ($\beta = 0.05$).}
\end{figure}

 With this choice of the 
parameters the map generates a signal with rapidly decaying correlations.
The short -- term
correlations are reflected in the fact that the spectrum is a white noise
for frequencies smaller than $\sim 0.1$, as shown in Fig. \ref{intermittency2}a.
The same chaotic system in the presence of additive noise is considered in Figures \ref{intermittency2}b and 
\ref{intermittency2}c.
The noise values are independently
drawn from a Gaussian distribution with standard deviation of $0.4$ (the standard deviation
of the clean signal is $0.26$). 
The perturbed chaotic signal enters as input in the KNNR
algorithm.
In Fig. \ref{intermittency1} is shown how
the KNNR algorithm is capable to reduce
considerabily the noise. Morover, the filtered signal has similar spectral properties that
the clean signal, despite the fact that the noisy data displays an almost flat spectrum at all 
frequencies.

\begin{figure}[h] 
\vskip0.5cm
\centering{\resizebox{6cm}{!}{\includegraphics{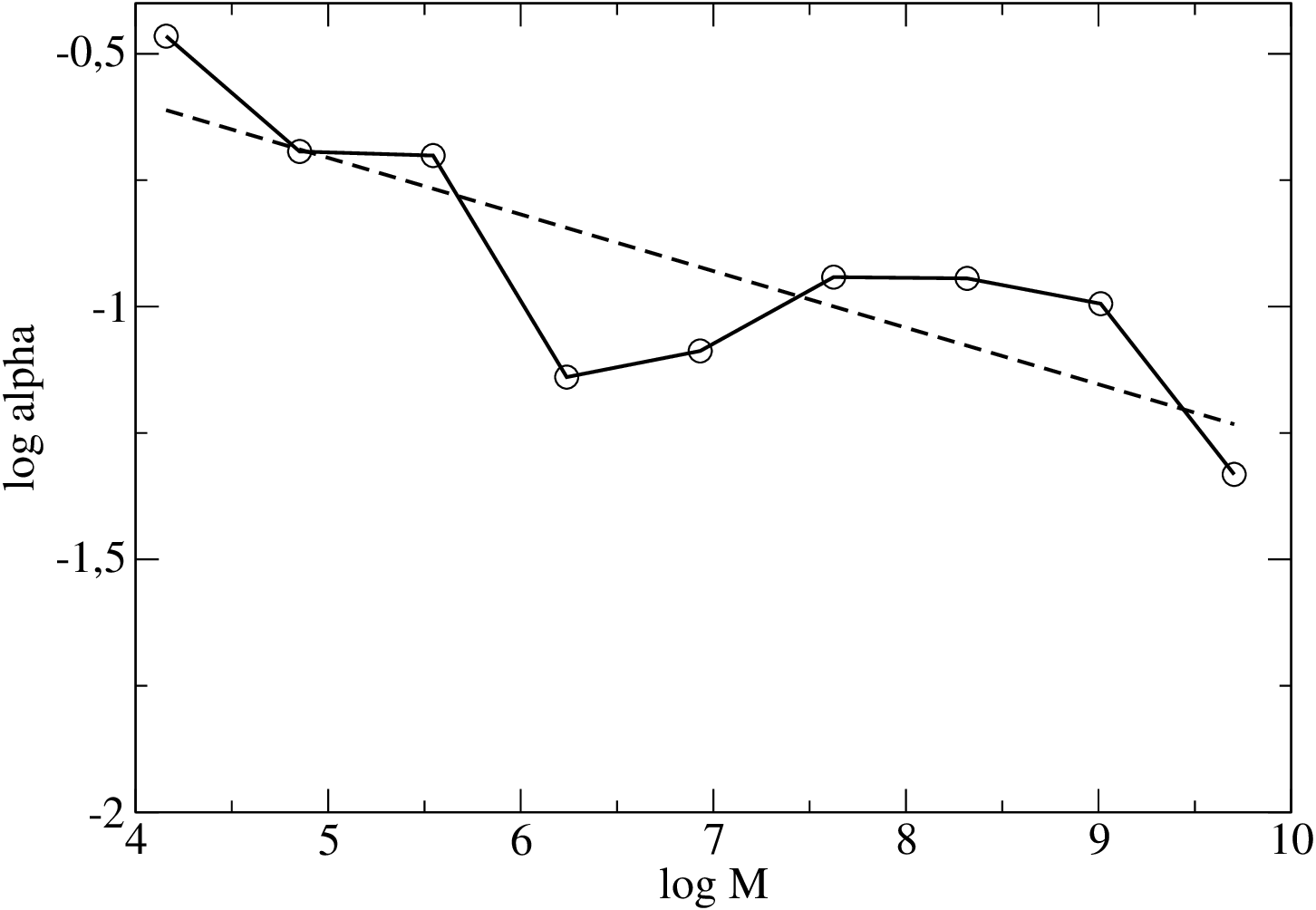}}}
   \caption{\label{alphaIntermittency} Behavior of $\alpha$ with increasing $M$
for the noisy intermittency map ($\beta = 0.05$).}
\end{figure}

The behavior of $\alpha$ calculated from samples
of the noisy signal with increasing sample size is shown in Fig. \ref{alphaIntermittency}. 
The $F$--test gives $F = 12.8 > F_{0.05}(1,7) = 5.59$.

 ii) $\beta = 0.0005$

 In this case the correlations decay much more slowly. The mean
squares fit of the power
spectrum of the clean signal to a power law
indicates $P(f) \propto f^{-1.15}$, with a crossover to white noise at frequencies $\sim 0.001$.

\begin{figure}[h] 
\vskip0.5cm
\centering{\resizebox{6cm}{!}{\includegraphics{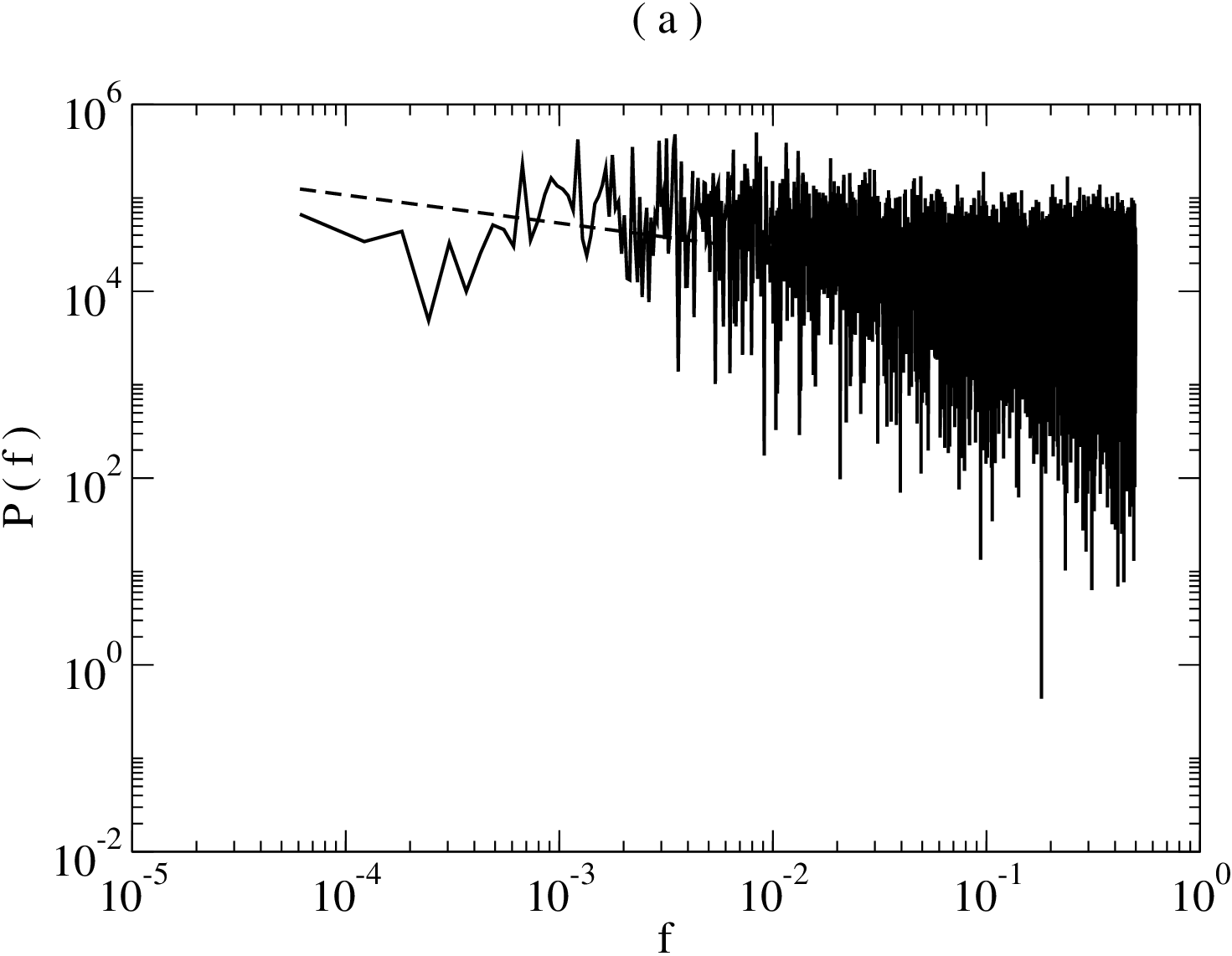}}}
\vskip1.0cm
\centering{\resizebox{6cm}{!}{\includegraphics{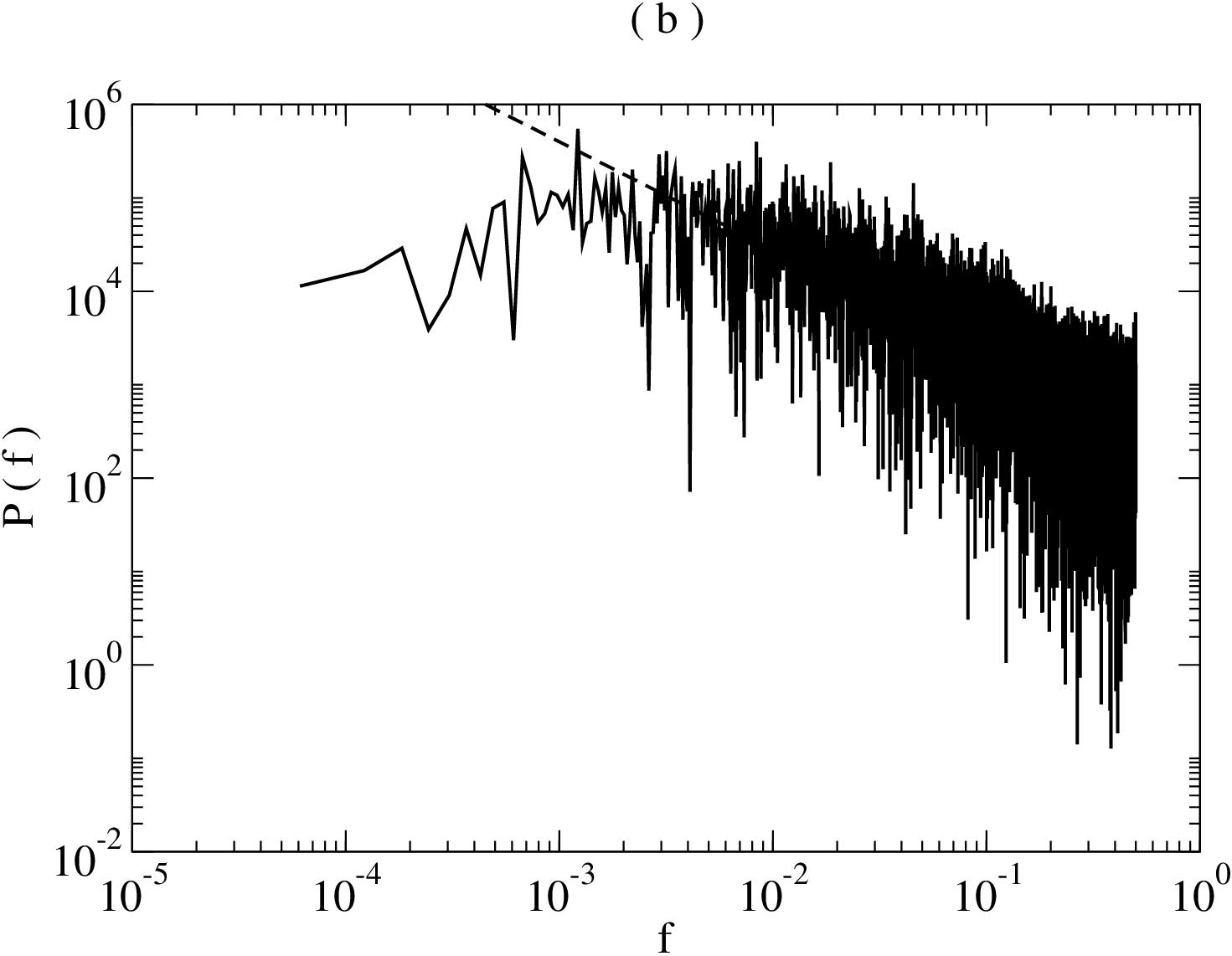}}}
\vskip1.0cm
\centering{\resizebox{6cm}{!}{\includegraphics{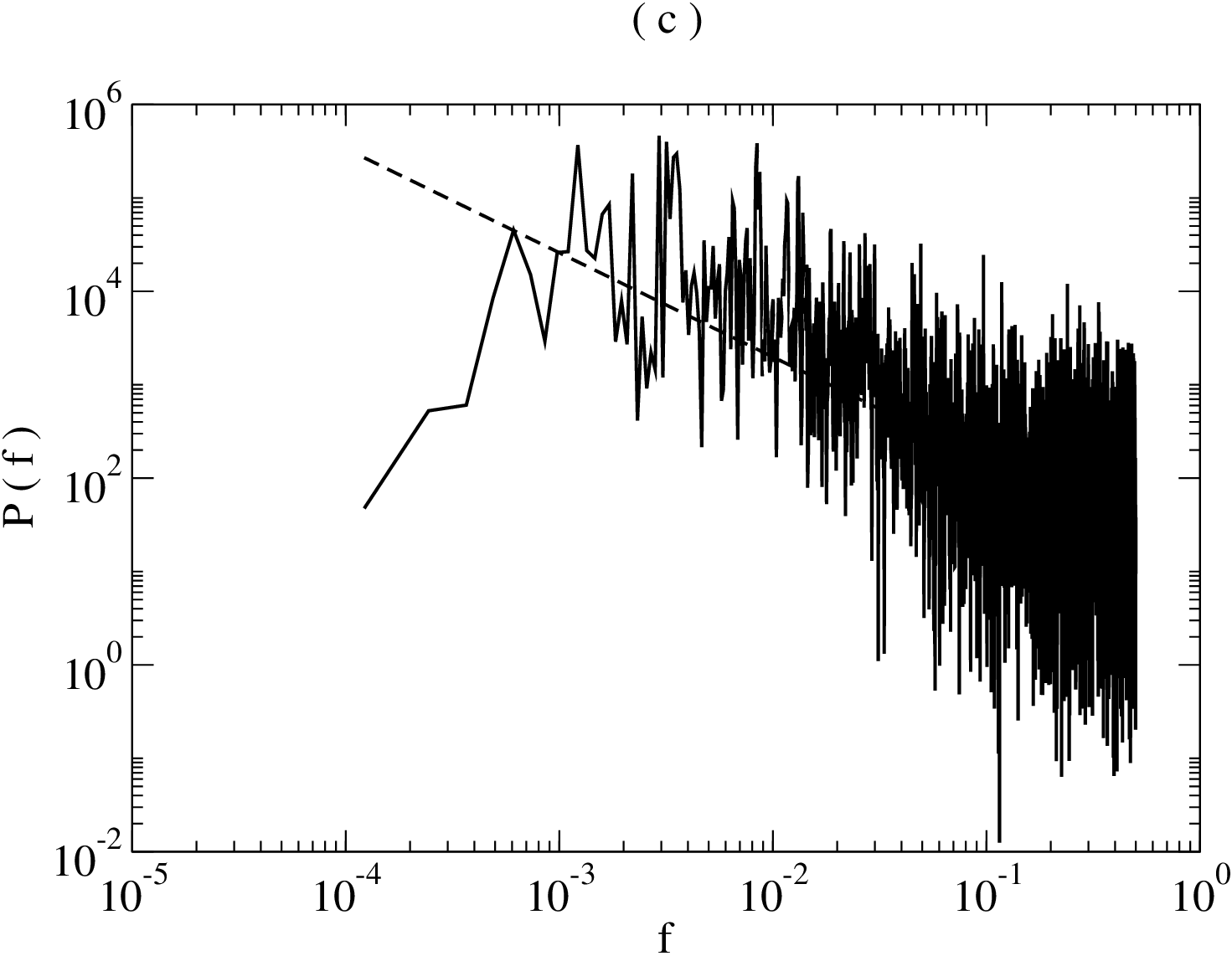}}}
   \caption{\label{intermittency3} Log--log plots of the power spectra of a signal
generated by the intermittency map ($\beta = 0.0005$): (a) noisy case, (b) clean signal, 
(c) filtered signal.
The clean and filtered signals display very similar spectral properties,
while the noisy signal is close to a white noise.}
\end{figure}

 Noise reduction is performed to this map in the presence of independent
Gaussian perturbations, taken from a distribution with standard deviation of
$0.5$, a value that almost doubles the standard deviation of the clean signal, which 
is $0.26$.
The Fig. \ref{intermittency3} makes clear how the KNNR algorithm is in this
case capable
to extract the essentially correct spectral properties from a very noisy input signal.
While the noisy signal has a power spectrum described by $P(f) \propto f ^{-0.3}$,
which is close to the spectrum of
a white noise, the fitting of the spectrum of the filtred signal to a power law
indicates $P(f) \propto f ^{-1.11}$.

The application of the $F$--test to succesive values of $\alpha$ calculated by
the KNNR algorithm with the noisy signal as input, 
gives evidence of the convergence to a finite $L$.
It is found $F = 13 > F_{0.05}(1,7) = 5.59$.

\subsection{White Noise and Ornstein--Uhlenbeck Processes} \label{stochastic}

In contrast to deterministic systems, even in the case that these were chaotic,
stochastic systems do not display a convergence in $L$ with increasing observation time.
The numerical experiments indicate that 
for signals generated by stochastic processes with a finite correlation length,
$L$ asimptotically grows linearly with sample size.

 The KNNR algorithm is applied to signals generated by discrete
analogous of the white noise and Ornstein--Uhlenbeck processes: sequences of
independent random numbers and the AR(1) process, respectively.

\begin{figure}[h] 
\vskip0.5cm
\centering{\resizebox{6cm}{!}{\includegraphics{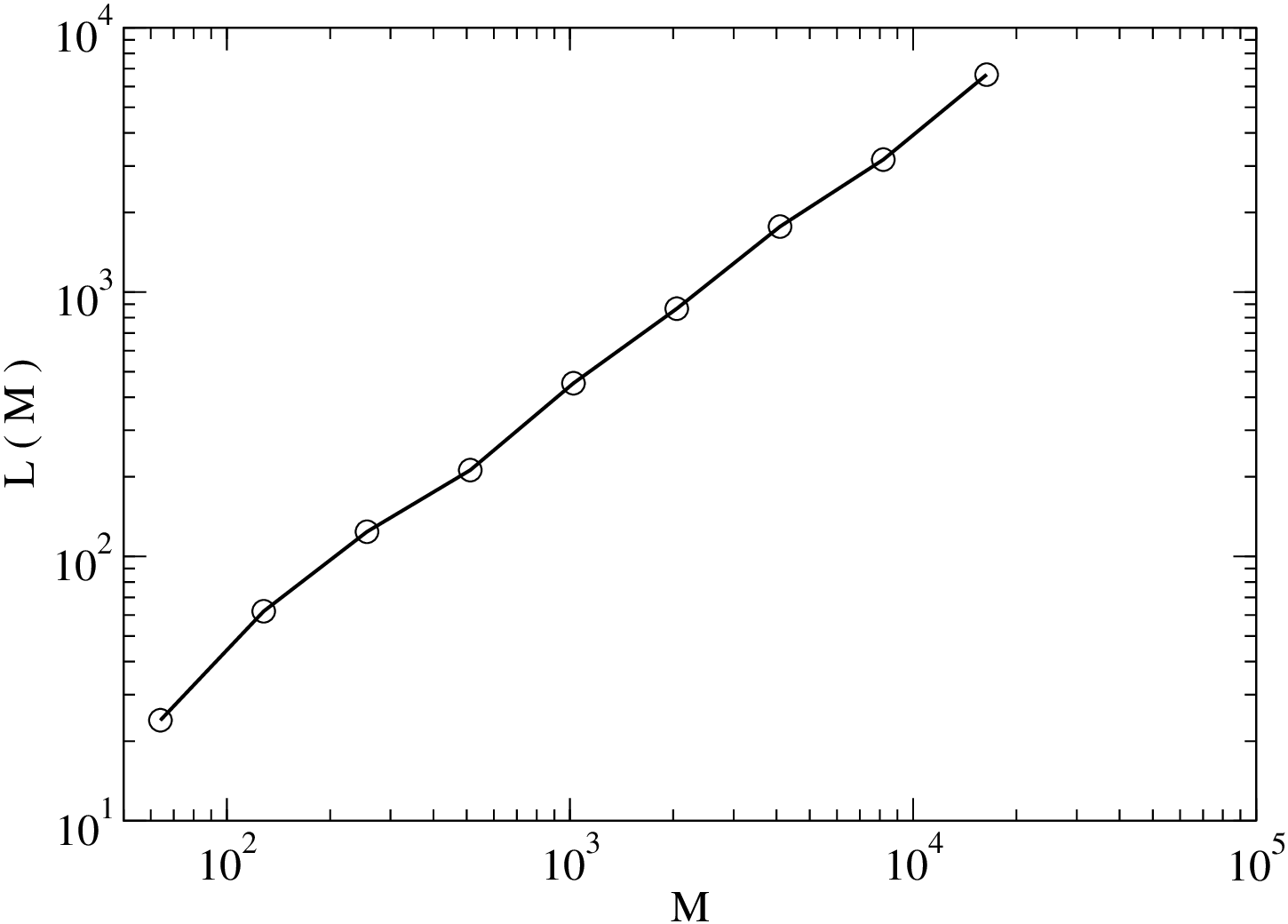}}}
   \caption{\label{divergenceWN} $L$ $vs$. $M$ for  a sequence of independent random numbers.
$L$ diverges linearly with sample size. }
\end{figure}

\begin{figure}[h] 
\vskip0.5cm
\centering{\resizebox{6cm}{!}{\includegraphics{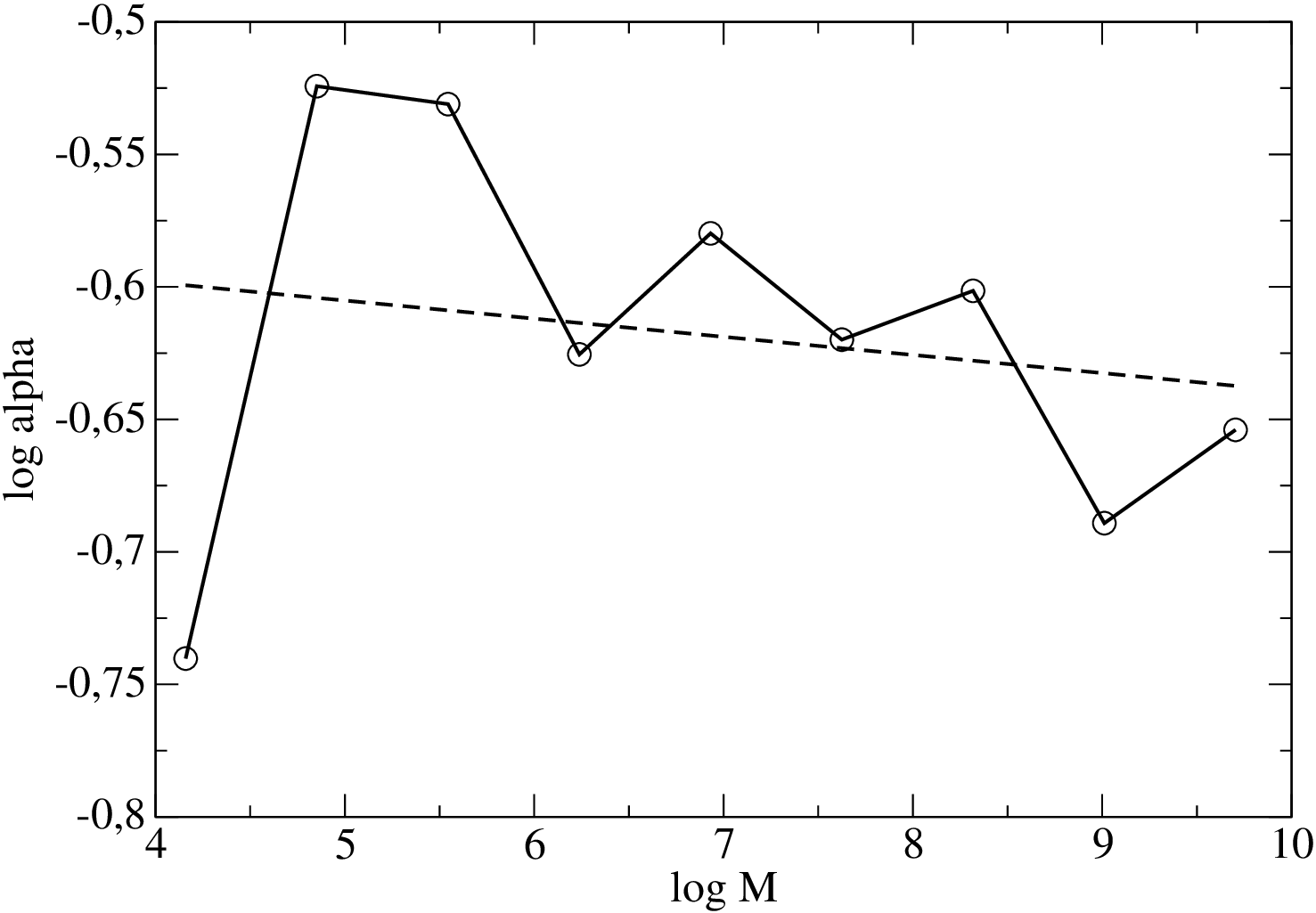}}}
   \caption{\label{alphaWn} Behavior of $\alpha$ with increasing $M$
for a sequence of independent random numbers.}
\end{figure}

 A sequence of independent 
Gaussian deviates is generated by the already mentioned L'Ecuyer
algorithm.
In Fig. \ref{divergenceWN} is presented
the behavior of model complexity for a signal in which the random numbers are
drawn from a distribution with standard deviation of $0.23$. The number $L$ 
diverge linearly. 
Performing an $F$-test in the same way as before (Fig. \ref{alphaWn}) gives 
$F = 0.25 << F_{0.05}(1,7) = 5.59$, which indicates that the hypothesis
of a constant $\alpha$ can't be rejected at the $95 \%$ confidence level.

\begin{figure}[h] 
\vskip0.5cm
\centering{\resizebox{6cm}{!}{\includegraphics{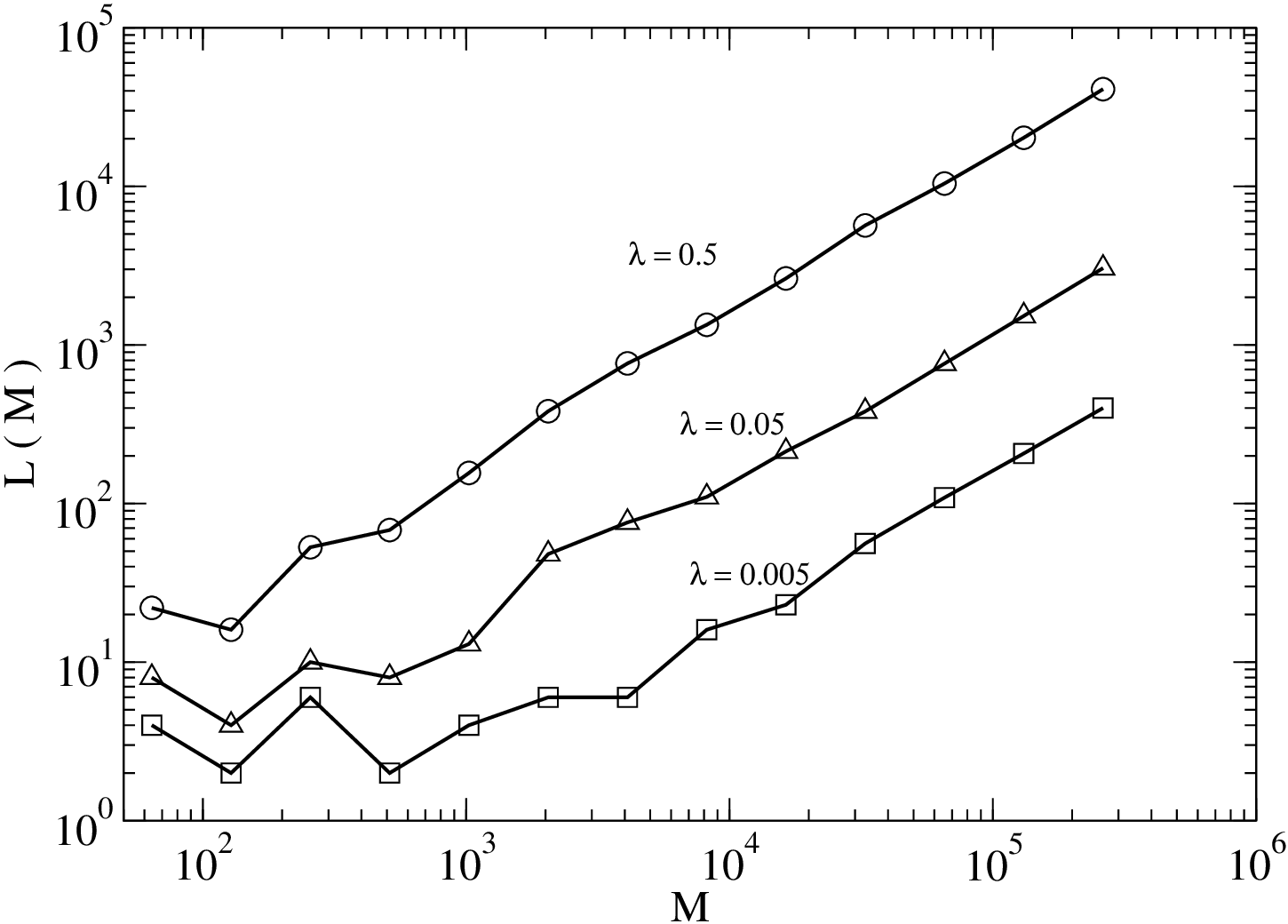}}}
   \caption{\label{divergenceOU} Behavior of $L$ 
with increasing $M$ for AR(1) processes with
different correlation lengths.}
\end{figure}

 In Fig. \ref{divergenceOU} the values of $L$ for increasing sample size
are plotted for three different realizations of an AR(1) process of the form

\begin{eqnarray}\label{ou}
y_{t}=y_{t-1} - \lambda y_{t} + \varepsilon_{t}, \quad 0 < \lambda < 1 .
\end{eqnarray}

 The term $\varepsilon_{t}$ is again a Gaussian deviate generated by the
L'Ecuyer algorithm with a standard deviation of $0.23$. 
In the limit in which the parameter $\lambda$ goes to zero the
process (\ref{ou}) tends to a random walk. For other values of $\lambda$ the correlations
decay exponentially, with a characteristic time $1/\lambda$. 
The examples considered in Fig. \ref{divergenceOU} 
have the parameter values $\lambda = 0.5$, $\lambda = 0.05$ and
$\lambda = 0.005$. Note that $L$ eventually diverges linearly for all cases. 
The point at which
this divergent regime is attained depends on the correlation length.
In fact, the $F$--test performed for a maximum sample size of $16384$ rejects the 
null hypothesis at a $95 \% $ confidence level only in the first two cases. 
This implies that for small enough samples, linear autocorrelated
stochastic processes are indistinguishable 
by the KNNR algorithm
from chaotic sistems with similar 
autocorrelations.
This is quite natural taking into account
that the KNNR algorithm is based on the linear autocorrelation structure.
It must be pointed out however, that if the stochastic signal at hands
has finite correlation length, the numerical experiments suggest that the 
identification of determinism is always possible with large enough sample
sizes. 

It's worth mention that in all of the examples in this Subsection, the filtered signals
display the same correlation lengths than the original signals.

\section{Conclusion}\label{conclusion}

The proposed formalism constitute a basis
for a novel technique of identification of deterministic
behavior in time series.
A careful study of the
convergence of $L$ 
as the sample size grows, may 
be used to improve the introduced
statistical test. The question
of the definition of the most adequate statistic and test to be used,
e. g. parametric or non--parametric, deserves further research.
In the same direction, statistical tests could also
be made on the basis of a
comparision between the spectral properties of a signal
before and after its
filtering by the KNNR algorithm. 

 The presented results, on the other hand, give
a linear filter for noise reduction capable to extract features
otherwise difficult to deduce from traditional linear approches.
Further research should be done on the use of the 
KNNR algorithm for noise reduction and forecasting
in important fields of application.

 The presented approach treats the
time series in a very direct manner. 
The generalization of the KNNR algorithm to the case
in which $y$ depends on more than one variable could be used to allow
delay representations of data. This may give a more powerful algorithm, 
capable to identify deterministic behavior in smaller data sets, and to connect
the presented theory with the important problem of 
the calculation of embedding dimensions. This generalization of the study to
higher dimensional data sets could also find application in questions such like
the estimation of the optimal number of hidden neurons in 
models of artificial neural
networks. 

\section*{Acknowledgment}
The author would like to thank to CONACYT, 
SEP--PROMEP and UANL--PAICYT for
partial financial support.


\begin{thebibliography}{1}


\bibitem{berrones}F. ~Bagnoli, A. ~Berrones and F. ~Franci, ``Degustibus Disputandum
(Forecasting Opinions by Knowledge Networks)'', \emph  {Physica A} {\bf 332}, 
2004, pp. 509-518.

\bibitem{barahona} M. ~Barahona and C. ~Poon, `` Detection of Nonlinear Dynamics
in Short, Noisy Time Series: '', 
\emph  {Nature} {\bf 381}, 1996, pp. 215-217.

\bibitem{encBerrones} A. ~Berrones, ``Filtering by Sparsely Connected 
Networks Under the Presence of Strong Additive Noise'', 
to be published in \emph  {Proc. Seventh Mexican International Conference on
Computer Science (ENC06)}.

\bibitem{haykin} S. ~Haykin, \emph {Neural Networks: a Comprehensive Foundation}, Prentice Hall,
1999.

\bibitem{henon} M. H\'enon, ``A Two--Dimensional Mapping with a Strange
Attractor'', 
\emph  {Commun. Math. Phys.} {\bf 50}, 1976, 69.

\bibitem{herramilli} A. ~Herramilli, R. ~Singh and P. ~Pruthi
``Chaotic Maps as Models of Packet Traffic'', 
\emph  {Proc. 14th Int. Teletraffic Cong.} {\bf 35}, Elsevier, 1994.

\bibitem{kantz} H. ~Kantz and T. ~Schreiber, \emph {Nonlinear Time Series Analysis}.
\hskip 1em plus 0.5em minus 0.4em\relax
Cambridge University Press, 2004.

\bibitem{maslov} S. ~Maslov and Y. ~Zhang, ``Extracting Hidden Information from
Knowledge Networks'', \emph {Physical Review Letters} {\bf 87}, 2001, 248701.

\bibitem{maslov2} S. ~Maslov and K. ~Sneppen, ``Specificity and Stability in Topology
of Protein Networks'', 
\emph {Science} {\bf 296}, 2002, 910.

\bibitem{ott}
E. ~Ott, \emph{Chaos in Dynamical Systems}. 
\hskip 1em plus 0.5em minus 0.4em\relax
Cambridge University Press, 1993.

\bibitem{nr}
W. ~H. Press, S. ~A. Teukolsky, W. ~T. Vetterling and B. ~P. Flannery,
\emph{Numerical Recipes in C++. The Art of Scientific Computing}, 2nd~ed.
\hskip 1em plus 0.5em minus 0.4em\relax
Cambridge University Press, 2002.

\bibitem{renaud} O. ~Renaud, J. ~Starck and F. ~Murtagh, 
``Wavelet--Based Combined Signal Filtering and Prediction'', 
\emph  {IEEE Transactions on Systems, Man and Cybernetics--Part B} {\bf 35}, 2005,
pp. 1241-1251.

\bibitem{shannon} C. E. ~Shannon and W. ~Weaver, \emph{The Mathematical Theory of
Information}, 
\hskip 1em plus 0.5em minus 0.4em\relax
University of Illinois Press, 1949.

\bibitem{schuster} H. G. ~Schuster \emph{Deterministic Chaos: An Introduction},
2nd revised ~ed. 
\hskip 1em plus 0.5em minus 0.4em\relax
VCH, 1988.

\bibitem{surrogate} J. ~Theiler, S. ~Eubank, A. ~Longtin, B. ~Galdrikian,
and J. D. ~Farmer, ``Testing for Nonlinearity in Time Series: the Method of
Surrogate Data'', \emph{Physica D} {\bf 58}, 1992, pp. 77-94.

\bibitem{wiener} N. ~Wiener, \emph{Cybernetics: Or the Control and
Communication in the Animal and the Machine}, 2nd~ed.
\hskip 1em plus 0.5em minus 0.4em\relax
MIT Press, 1965.


\end{thebibliography}
\end{document}